\documentclass[aps,showpacs,preprintnumbers,amsmath, amssymb]{revtex4}

\UseRawInputEncoding
\usepackage{float}
\usepackage{graphics,epsfig,hyperref}
\usepackage{graphicx}
\usepackage{dcolumn}
\usepackage{bm}
\usepackage{epstopdf}
\usepackage{subfigure}

\def \be {\begin{equation}}
\def \ee {\end{equation}}
\def \bea {\begin{eqnarray}}
\def \eea {\end{eqnarray}}

\begin{document}
\baselineskip=0.8 cm
\title{\bf Chaotic motion of scalar particle coupling to Chern-Simons invariant in the stationary axisymmetric Einstein-Maxwell dilaton black hole spacetime}
\author{Lina Zhang$^{1,2}$, Songbai Chen$^{1,2,3}$\footnote{csb3752@hunnu.edu.cn}, Qiyuan Pan$^{1,2,3}$\footnote{panqiyuan@hunnu.edu.cn}, and Jiliang Jing$^{1,2,3}$ \footnote{jljing@hunnu.edu.cn}}

\affiliation{$^{1}$Key Laboratory of Low Dimensional Quantum Structures and Quantum Control of Ministry of Education, Synergetic Innovation Center for Quantum Effects and Applications, and Department of Physics, Hunan Normal University, Changsha, Hunan
410081, China}
\affiliation{$^{2}$Institute of Interdisciplinary Studies, Hunan Normal University, Changsha, Hunan 410081, China}

\affiliation{$^{3}$Center for Gravitation and Cosmology, College of Physical Science and Technology, Yangzhou University, Yangzhou 225009, China}

\begin{abstract}
\baselineskip=0.6 cm
\begin{center}
{\bf Abstract}
\end{center}

We investigate the motion of a test scalar particle coupling to the Chern-Simons (CS) invariant in the background of a stationary axisymmetric black hole in the Einstein-Maxwell-Dilaton-Axion (EMDA) gravity. Comparing with the case of a Kerr black hole, we observe that the presence of the dilation parameter makes the CS invariant more complex, and changes the range of the coupling parameter and the spin parameter where the chaotic motion appears for the scalar particle. Moreover, we find that the coupling parameter together with the spin parameter also affects the range of the dilation parameter where the chaos occurs. We also probe the effects of the dilation parameter on the chaotic strength of the chaotic orbits for the coupled particle. Our results indicate that the coupling between the CS invariant and the scalar particle  yields the richer dynamical behavior of the particle in the rotating EMDA black hole spacetime.

\end{abstract}

\maketitle
\newpage

\section{Introduction}

Chaos is a kind of very interesting motions occurred in nonlinear dynamical systems. The main feature of chaos is its high sensitivity to initial conditions, so tiny differences in initial conditions grow at exponential rates and lead to totally different states of motion \cite{Ott,Brown,Brown1}. This means that a long-term prediction to the motion is very difficult for a chaotic dynamical system. Therefore, the effects from nonlinear interactions yield that
chaotic systems have a lot of  novel properties, which are not shared by the
linear dynamical systems. This also triggers  much efforts being devoted to studies of chaos in
various physical fields.

In general relativity, to probe the chaotic motions of particles, one must resort to some spacetimes with complex
geometrical structures or introduce some extra interactions to ensure that the dynamical system
of particles is non-integrable. In this way, the chaotic orbits of particles have been investigated in the multi-black hole spacetime \cite{Dettmann}, or in a black hole spacetime immersed in a magnetic field \cite{Ldan}, or in an accelerating and rotating black hole spacetime \cite{schen1}. Recently, by introducing an extra interaction with the Einstein tensor, the chaotic dynamics of a scalar test particle in the Schwarzschild-Melvin black hole spacetime has been studied \cite{mschen2}.  Additionally, the thermal chaos in the extended phase space has been studied in the spacetimes of
a charged AdS black hole in various theories of gravity \cite{Chabab,ChenLi,Mahish,Dai}.

Although Einstein's general theory of relativity has passed all current observational and experimental tests \cite{Will}, it is widely recognized that it may not be the ultimate theory describing gravitational fields. Instead, it could be a valid description of an unknown fundamental theory of gravity \cite{ Berti}. Therefore,  there is great interest in studying potential extensions to Einstein's general relativity. One of the most promising alternative gravity theories is the dynamical CS modified gravity \cite{Alexander}, where the Einstein-Hilbert action is modified by adding an extra interaction between the scalar field and the CS invariant. This interaction captures the leading-order gravitational parity violation. Generally, it is not easy to get an analytical black hole solution in the dynamical CS gravity because both motion equations of gravitational and scalar fields must be satisfied simultaneously. Thus, the analytical solution of a rotating black hole in this modified gravity has been obtained only in the small-coupling and/or slow-rotation limit \cite{Shiromizu,N. Yunes, K. Konno, K. Yagi,Nashed,G. Nashed}. Recently, a CS scalar field induced by a rapidly rotating black hole in the dynamical CS modified gravity has been also investigated in \cite{Konno}, and it is shown that the scalar field diverges on the inner horizon although it is regular on the outer horizon and vanishes at infinity, which means that the CS scalar field becomes problematic on the inner horizon.

Only the linear coupling is discussed in above literatures on the dynamical CS gravity, where the coupling term with the CS invariant is proportional to the scalar field. Inspired by the theoretical model in the quadratic scalar-Gauss-Bonnet gravity \cite{ Antoniou, Doneva, Silva}, the model where the CS invariant is coupled to the quadratic function of the dynamical scalar field has been investigated in \cite{Yuan}, and it is found that the scalar perturbation around a Kerr black hole grows at exponential rate in a certain regions of the parameter space, which means that the black hole could be unstable under such CS scalar perturbation. With the short wave approximation study, Zhou et al. \cite{Zhou} studied the effects of such quadratic coupling on the motion of a test scalar particle in a Kerr black hole background, and found that there exists the chaotic phenomenon in the motion of the scalar particle coupling to the CS invariant. It is natural to ask whether there exists the chaos in the motion of the scalar particle in other rotating black hole spacetimes under such kind of couplings. The EMDA black hole is an important black hole in the EMDA gravity, which is regraded as the low energy limit of the heterotic string theory \cite{Kerrsen}. This black hole is characterized by its mass, spin and dilation parameters.
Since the string theory is the present strongest candidate for the quantum description of gravity, the EMDA black hole has been widely studied in various aspects. In this paper, we want to study the effects of such quadratic coupling
on the motion of a test scalar particle in a EMDA black hole background \cite{Kerrsen} and probe effects of the coupling parameter together dilation parameter on the motion of the scalar particle coupling to the CS invariant.

The paper is organized as follows. In Sec. 2, we adopt the short-wave approximation as in \cite{Zhou} and present the geodesic equation of a test scalar particle coupling to the CS invariant in the EMDA black hole spacetime. In Sec. 3, we investigate the
chaotic motion of the coupled scalar particle with the Poincar\'{e} section, the fast Lyapunov indicator (FLI), the bifurcation
diagram and the basins of attraction. We probe the effects of this coupling together with the black hole dilation and spin parameters on the chaotic behavior of the coupled scalar particle. Finally, we end the paper with a summary.

\section{Geodesics of scalar particle coupling to Chern-Simons invariant in the stationary axisymmetric EMDA black hole}

In the theory of Einstein-Maxwell dilation gravity, the action containing the coupling between the CS invariant and
the quadratic function of a scalar perturbational field can be expressed as \cite{Yuan}
\begin{equation}
S=\int d^{4} x \sqrt{-g}\left(\frac{R}{16\pi G}-\frac{1}{2} g^{\mu\nu}\partial_{\mu} \phi \partial^{\mu} \phi-\frac{1}{4}e^{-2\phi}F^{\mu\nu}F_{\mu\nu}
-\frac{1}{2}g^{\mu\nu} \partial_{\mu} \Phi \partial^{\mu} \Phi-\mu^2 \Phi^2+\alpha\;  ^{*}R R \Phi^2\right),\label{act1}
\end{equation}
where $R$ is the Ricci scalar and $G$ is the usual Newton's constant. $\phi$ is a massless dilaton field and $\Phi$ is a massive scalar perturbational field with the mass $\mu$.  The CS invariant $^{*} R R$ is defined by
\begin{equation}
^{*} R R=\frac{1}{2} \epsilon^{\alpha \beta \gamma \delta} R_{\;\nu \gamma \delta}^{\mu} R_{\;\mu \alpha \beta}^{\nu},
\end{equation}
which is a topological invariant.  $\epsilon^{\alpha \beta \gamma \delta}$ is the four-dimensional Levi-Civit\`{a}
tensor and $R_{\;\nu \gamma \delta}^{\mu} $ is the Riemann curvature tensor. The coupling parameter $\alpha$ has a dimension of length squared and describes the strength of the coupling between the scalar perturbational field and the CS invariant.
Varying the action (\ref{act1})  with respect to the field $\Phi$, one can obtain the modified Klein-Gordon equation for the scalar perturbation
\begin{equation}
\frac{1}{\sqrt{-g}}\frac{\partial}{\partial x^{\mu}}\left(\sqrt{-g}g^{\mu\nu}\frac{\partial \Phi}{\partial x^{\nu}}\right)+(2 \alpha^{*} R R-\mu^2)\Phi=0.\label{cKGordon}
\end{equation}
As discussed in \cite{Zhou}, we here adopt the short wave approximation for the scalar perturbational field $\Phi$ and derive the equation of motion for scalar particle in a background spacetime. In the short wave approximation,  the wavelength
of the scalar perturbational field is assumed to be much smaller than the typical curvature of the spacetime, so the particle aspect of the scalar field is dominated and its wave aspect can be neglected.
With this approximation, the scalar perturbational field $\Phi$ can be further reduced to
\begin{equation}\label{shwave}
\Phi=fe^{iS},
\end{equation}
where the amplitude $f$ is a small slowly-varying real and the phase $S$ is rapidly changing.
Therefore, the derivative term $f_{;\mu}$ can be ignored because it is not dominated in this case. The wave vector $\partial_{\mu}S$ can be regarded as the momentum $p_{\mu}$ of the corresponding scalar particle. And then the modified Klein-Gordon equation (\ref{cKGordon}) can be rewritten as
\begin{equation}\label{sesan}
g^{\mu\nu}p_{\mu}p_{\nu}-2\alpha^{*} R R=-1.
\end{equation}
Setting the mass of the scalar particle be $\mu=1$, the equation (\ref{sesan}) can be rewritten as a form of Hamilton-Jacobi equation
\begin{equation}\label{yelebi}
\frac{\partial S}{\partial \tau}+\mathcal{H}\left(x^{\mu},\frac{\partial S}{\partial x^{\mu}}\right)=0,
\end{equation}
where $S=\frac{1}{2}\tau+x^{\mu}p_{\mu}$ and the corresponding Hamiltonian is
\begin{equation}\label{Hamin}
\mathcal{H}\left(x^{\mu},\frac{\partial S}{\partial x^{\mu}}\right)=\frac{1}{2}g^{\mu\nu}p_{\mu}p_{\nu}-\alpha ^{*} RR.
\end{equation}
Here $x^{\mu}$ and $p_{\mu}$, respectively represent the spacetime coordinates and  the canonical momentum of the scalar field. $\tau$ is an affine parameter along a curve. The term $-\alpha ^{*} RR$ can be treated as an extra potential from the interaction between the scalar particle and the CS invariant.

In the Boyer-Lindquist coordinates $(t,r,\theta,\varphi)$, the solution of the stationary axisymmetric EMDA black hole has a form \cite{Kerrsen,Garcia,J,Pan}
\begin{eqnarray}\label{metric}
d s^{2}&=&-\frac{\Sigma-a^{2} \sin ^{2} \theta}{\Delta} d t^{2}-\frac{2 a \sin ^{2} \theta}{\Delta}[(r^{2}-2 D r+a^{2} )-\Sigma]dtd\varphi +\frac{\Delta}{\Sigma} d r^{2}\nonumber \\&&
 + \Delta d \theta^{2} + \frac{\sin ^{2} \theta}{\Delta}[(r^{2}-2 D r+a^{2} )^{2} -\Sigma a^{2} \sin ^{2} \theta] d\varphi^{2},
\end{eqnarray}
with
\begin{eqnarray}
\Sigma = r^{2}-2 M r+a^{2} ,~~~~~~~~~
\Delta = r^{2}-2 D r+a^{2} \cos ^{2} \theta,
\end{eqnarray}
where $M$, $D$ and $a$ represent the mass, dilaton and angular momentum per unit mass of the black hole, respectively. The Arnowitt-Deser-Misner (ADM) mass of the black hole (\ref{metric}) is $M_{ADM}=M-D$.
The outer and inner horizons are located ad $r_{\pm}=M_{ADM}+D\pm\sqrt{(M_{ADM}+D)^{2}-a^{2}}$.
The CS invariant for the black hole (\ref{metric}) is
\begin{eqnarray}\label{CSRR1}
^* R R&=&\frac{32 a (M-D) \cos \theta (a^{2} \cos ^{2} \theta +2 D r -3 r^{2}) }{(a^{2} \cos ^{2} \theta-2 D r+r^{2})^{6}}\{ a^{2} \cos ^{2} \theta[2 D^{2} -3 D (M+3r) +9Mr] \nonumber \\&&
-2 D^{2} (Mr +r^{2} -a^{2}) +3 r^{2}(D M + D r -Mr) \}.
\end{eqnarray}
When $D=0$, the invariant reduces to that in the Kerr black hole.  As in the Kerr case, the CS invariant $^* R R$ (\ref{CSRR1}) also contains the linear term of $\cos\theta$, and it's asymmetrical to the equatorial plane.
The chaotic motion of the scalar particle coupling to the CS invariants in the Kerr spacetime has been investigated in \cite{Zhou}. Here, we will study the effects of the dilaton parameter $D$ on the motion of the coupled scalar particles.

From the Hamiltonian (\ref{Hamin}), one can get the geodesic equation for the coupled scalar particle
\begin{equation}\label{wfE1}
\dot{t}=\frac{{g}_{\varphi\varphi} E+{g}_{t \varphi} L_{z}}{{g}_{t \varphi}^{2}-{g}_{t t} {g}_{\varphi\varphi}},\quad\quad\quad
\dot{\varphi}=-\frac{{g}_{t \varphi} E+{g}_{t t} L_{z}}{{g}_{t \varphi}^{2}-{g}_{t t} {g}_{\varphi\varphi}},
\end{equation}
and
\begin{equation}\label{wfE2}
\ddot{r}=\frac{1}{2} g^{rr}\left(g_{t t, r} \dot{t}^{2}-g_{r r, r} \dot{r}^{2}+g_{\theta \theta, r} \dot{\theta}^{2}+g_{\varphi\varphi, r} \dot{\varphi}^{2}+2 g_{t \varphi, r} \dot{t} \dot{\varphi}-2 g_{rr, \theta} \dot{r} \dot{\theta}+2 \alpha \frac{\partial ^{*} R R}{\partial r}\right),
\end{equation}
\begin{equation}\label{wfE3}
\ddot{\theta}=\frac{1}{2} g^{\theta \theta}\left(g_{t t, \theta} \dot{t}^{2}+g_{r r, \theta} \dot{r}^{2}-g_{\theta \theta, \theta} \dot{\theta}^{2}+g_{\varphi\varphi, \theta} \dot{\varphi}^{2}+2 g_{t \varphi,\theta} \dot{t} \dot{\varphi}-2 g_{\theta \theta, r} \dot{r} \dot{\theta}+2 \alpha \frac{\partial^{*} R R}{\partial \theta}\right),
\end{equation}
where
\begin{equation}\label{shouheng}
E=-(g_{tt}\dot{t}+g_{t\varphi}\dot{\varphi}),\quad\quad\quad L_{z}=g_{\varphi\varphi}\dot{\varphi}+g_{t\varphi}\dot{t}.
\end{equation}
In addition, the motion of the coupled scalar particle also satisfies the constrain condition
\begin{equation}\label{Hcon}
h=g_{t t} \dot{t}^{2}+g_{r r} \dot{r}^{2}+g_{\theta \theta} \dot{\theta}^{2}+g_{\varphi\varphi} \dot{\varphi}^{2}+2 g_{t \varphi} \dot{t} \dot{\varphi}+1-2 \alpha^{*} R R=0.
\end{equation}
As in the Kerr case \cite{Zhou}, the CS invariant (\ref{CSRR1}) results in the differential equation (\ref{Hcon}) not being variable-separable. Therefore, the motion of the scalar particle may exhibit the chaotic behavior due to its interaction with the CS invariant. In the next section, we explore the effects of the dilaton parameter $D$ together with the coupling parameter $\alpha$ and the spin parameter $a$ on the motion of the coupled scalar particles in the EMDA black hole spacetime (\ref{metric}).

\section{ Chaotic motion of scalar particles coupling to Chern-Simons invariant in the stationary axisymmetric EMDA black hole}

We are now to probe the chaotic motion of scalar particles coupling to the CS invariant in the stationary axisymmetric EMDA  black hole spacetime. Chaos is highly sensitive to initial values, so the tiny error can yield enormous deviations as the dynamical system is in a chaotic state. To avoid the
pseudo-chaos arising from errors in numerical calculations,  we here adopt the corrected fifth-order Runge-Kutta method \cite{DZMa1,DZMa2} to solve differential equations (\ref{wfE1})-(\ref{wfE3}), which effectively ensures the
high precision because at every integration step the numerical
deviation is pulled back in a least-squares shortest path by correcting the
velocities ($\dot{r}$, $\dot{\theta}$).

It is well known that the motion of the particle is entirely determined by its initial conditions and the parameters
of the system. Generally, the choice for the parameters and initial conditions of the particle should be arbitrary in the allowed range in physics. For a convenience, we here set a regular orbit in the noncoupling case as the initial motion orbit of the particle and then probe the change of the disorder degree of the particle orbit with the
coupling, the spin and the dilaton parameters. Here, the selected regular orbit can be obtained by setting the parameters $\{$ $E=0.95$, $M_{ADM}=1$, $L=3.05M$, $a=0.31$, $ D=-0.23$ $\}$  and the initial conditions $\{$ $ r(0)=11.5$, $\dot{r}(0)=0$,  $\theta(0)=\frac{\pi}{2}$ $\}$.

\begin{figure}[ht]
\includegraphics[width=5.5cm ]{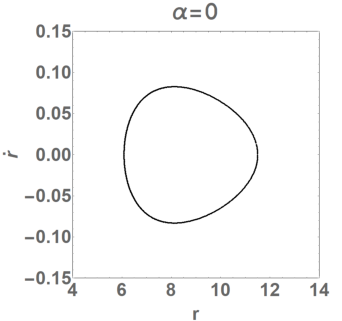}
\includegraphics[width=5.5cm ]{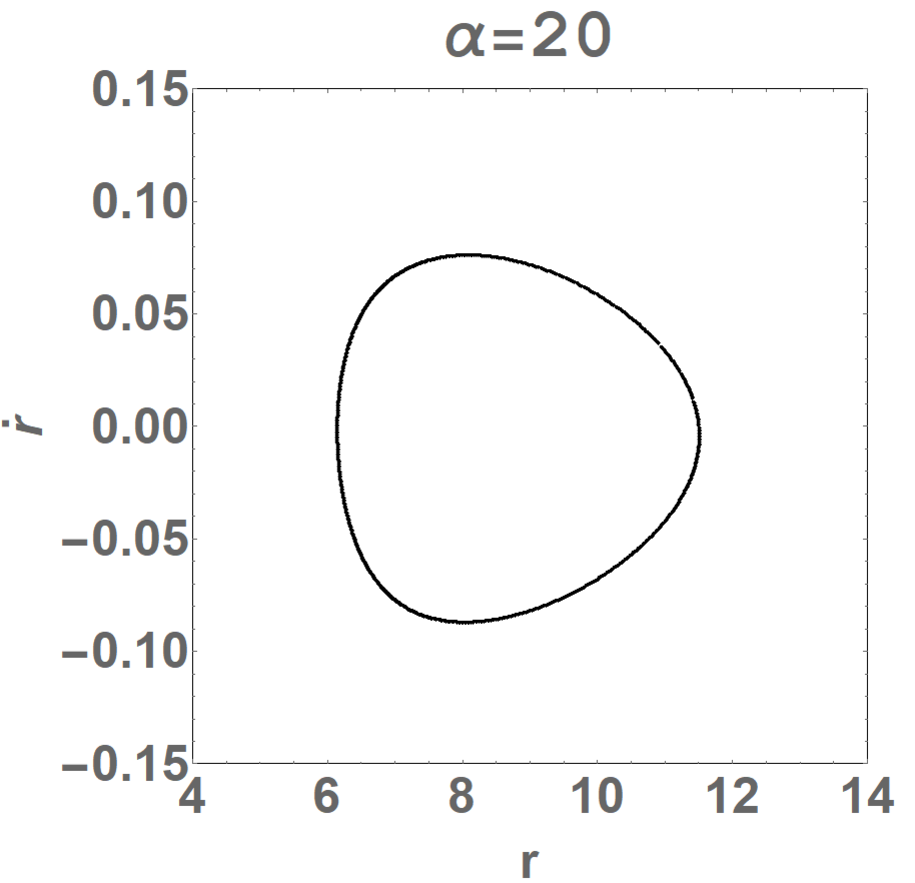}
\includegraphics[width=5.5cm ]{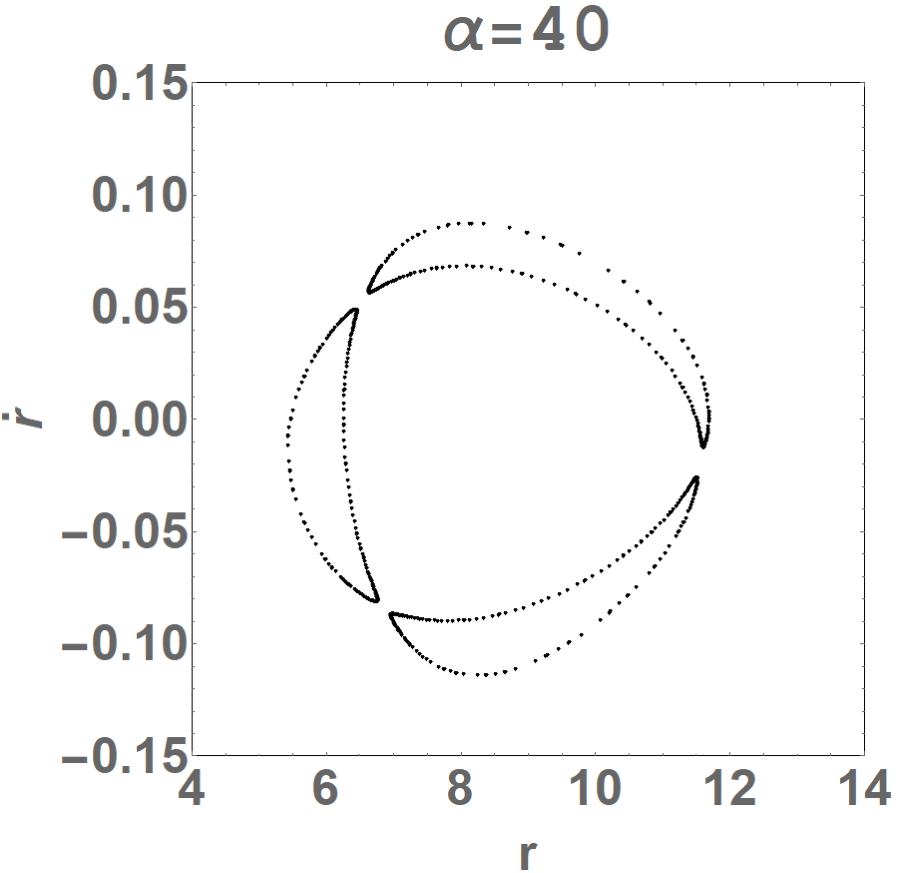}
\includegraphics[width=5.5cm ]{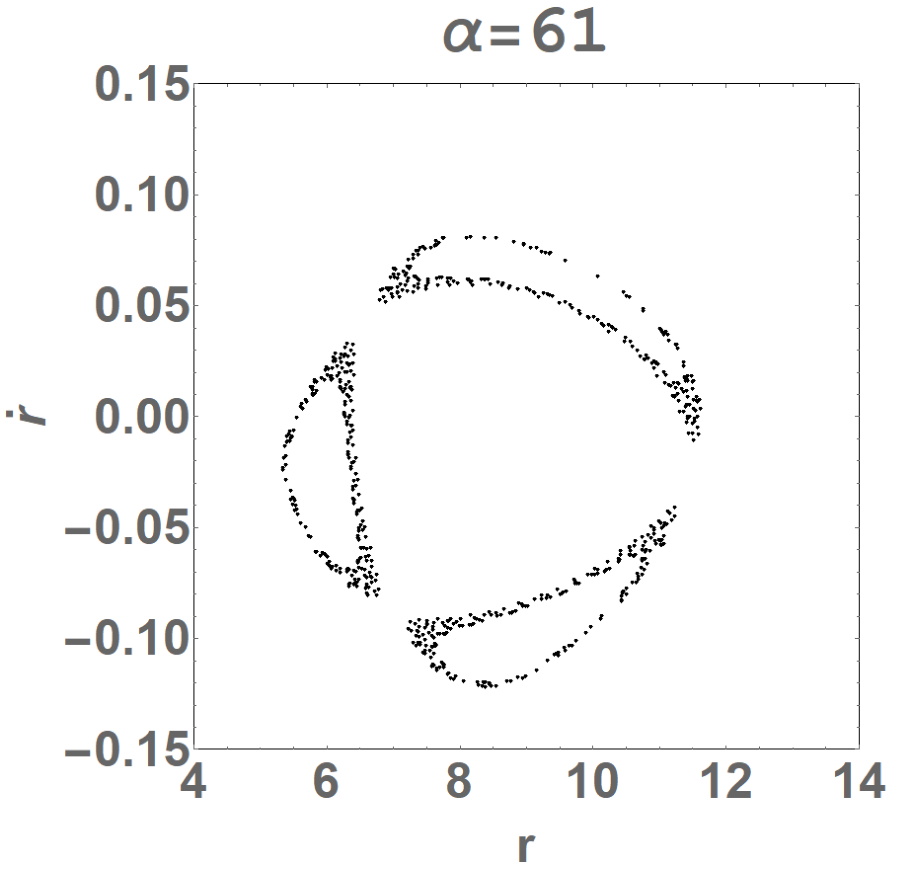}
\includegraphics[width=5.5cm ]{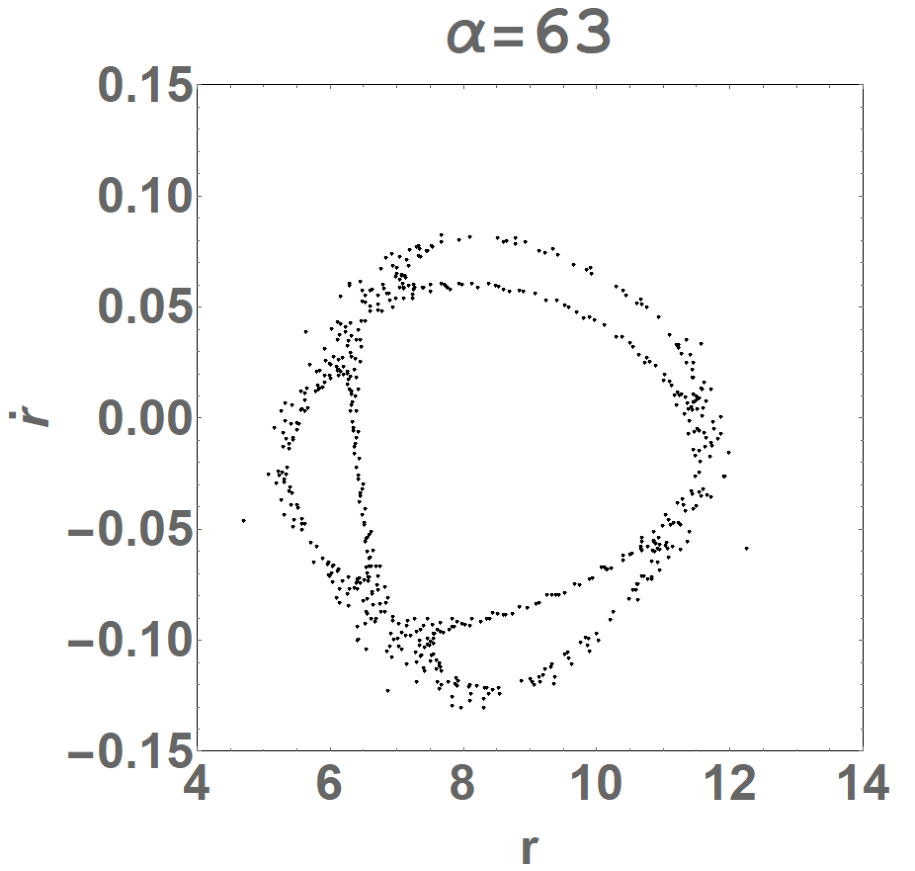}
\includegraphics[width=5.5cm ]{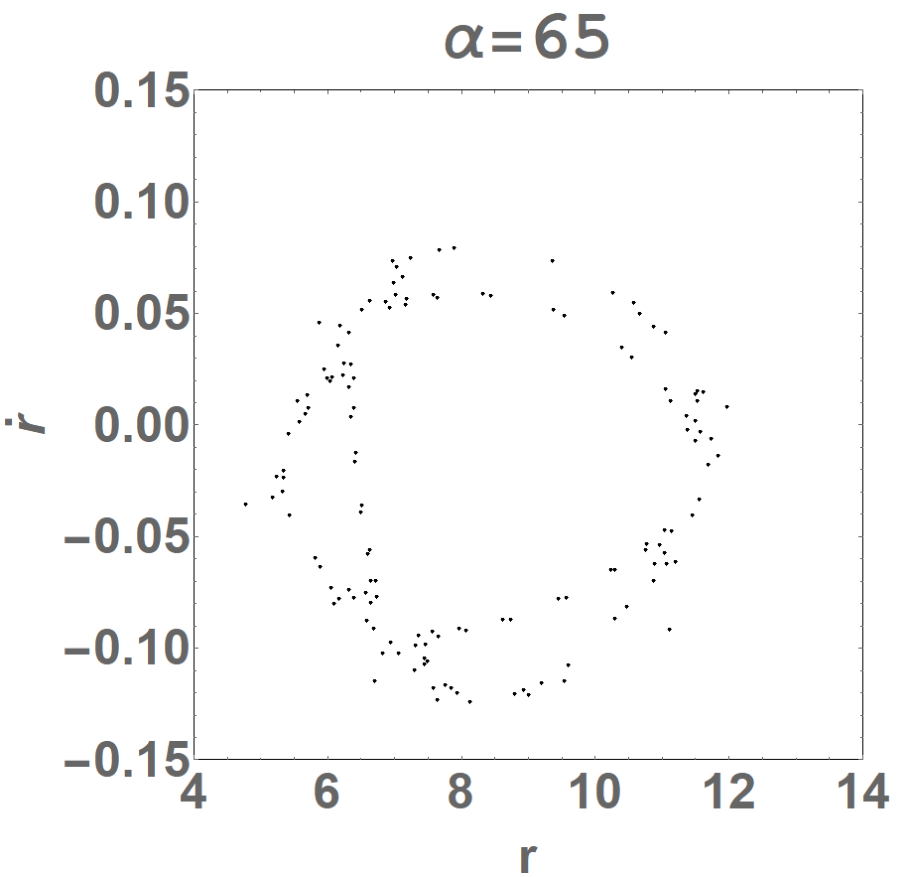}
\caption{The Poincar\'e section ($\theta = \frac{\pi}{2}$) with the coupling parameter $\alpha$ for the motion of a scalar particle coupling to the CS invariant in the stationary axisymmetric EMDA black hole for the fixed values of $r=11.5$, $a=0.31$, $D=-0.23$, $M_{ADM}=1$, $E=0.95$ and $L=3.05M$. }\label{fig1}
\end{figure}

Poincar\'{e} section is an effective method to discern the chaos because strange patterns of dispersed points with complex boundaries appear for the chaotic motion in the section. Fig. \ref{fig1} presents the changes of the Poincar\'e section ($\theta=\frac{\pi}{2}$) in the $r-\dot{r}$ plane with different CS coupling parameters $\alpha$. As $\alpha< 61$, we find that the phase path of the coupled scalar particle in the Poincar\'e section is a quasi-periodic Kolmogorov-Arnold-Moser (KAM) tori, which means that the motion of the particle is regular since its orbit moves on a torus in the phase space. Moreover, with the increase of $\alpha$, the KAM tori becomes progressively distorted. Especially, as $\alpha=40$, there is an island chain consisting of three secondary KAM toris, which belong to the same trajectory. When the CS coupling parameter is further increased to $\alpha=61$, one can find that the KAM tori is destroyed and many discrete points are randomly distributed in the section, which means that the motion of the particle is chaotic because its orbit is not limited to the original KAM torus as in regular motions. As $\alpha = 65$, we find that the number of discrete points in the Poincar\'e section decreases, which is caused by that the particles undergoing chaotic oscillations eventually fall into the black hole's event horizon or escapes to the spatial infinity. Therefore, the coupling of the CS invariant makes the motion of scalar particles more complex.

\begin{figure}[htbp!]
\includegraphics[width=5.5cm ]{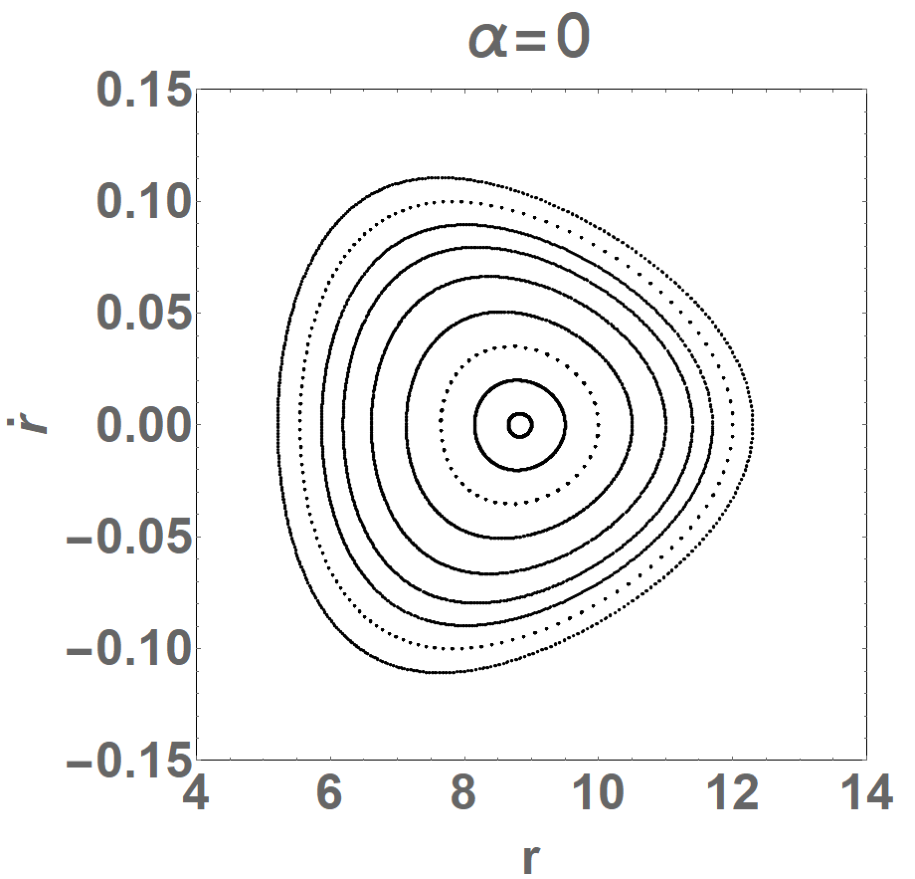}
\includegraphics[width=5.5cm ]{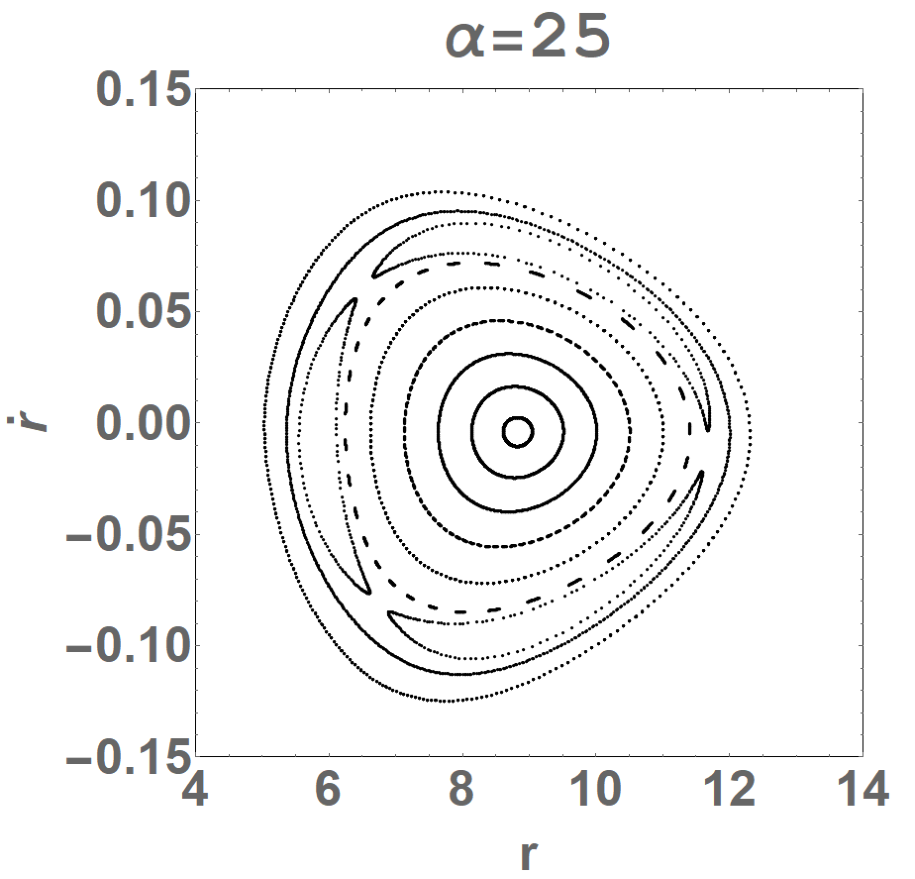}
\includegraphics[width=5.5cm ]{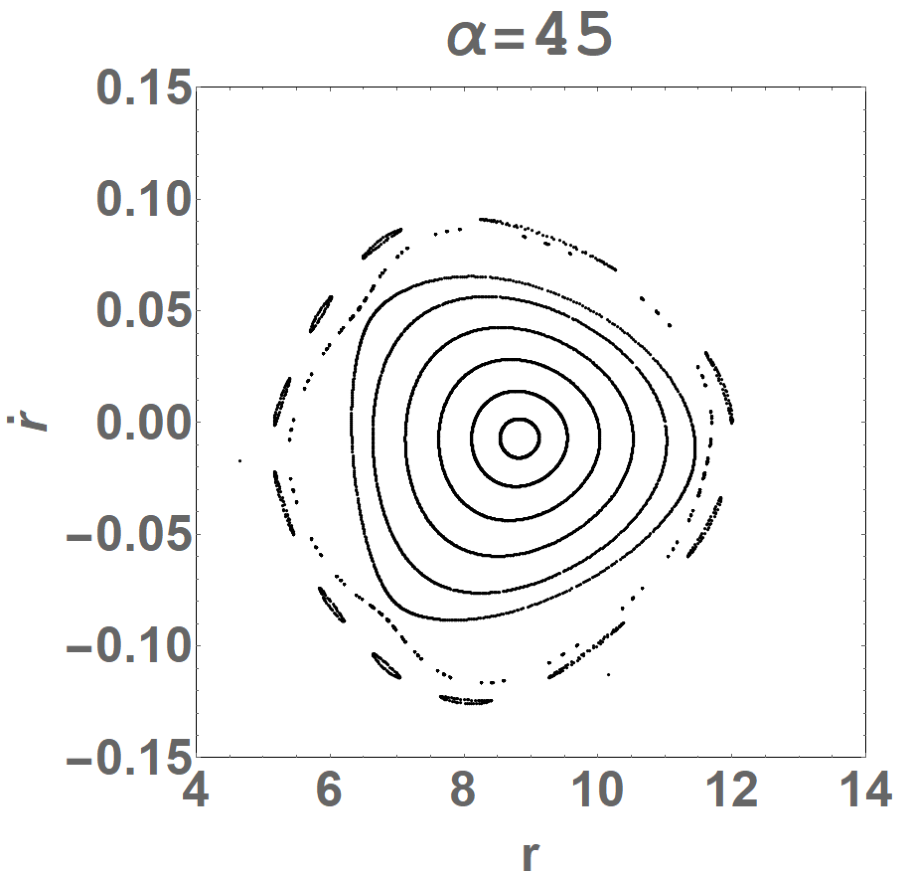}
\includegraphics[width=5.5cm ]{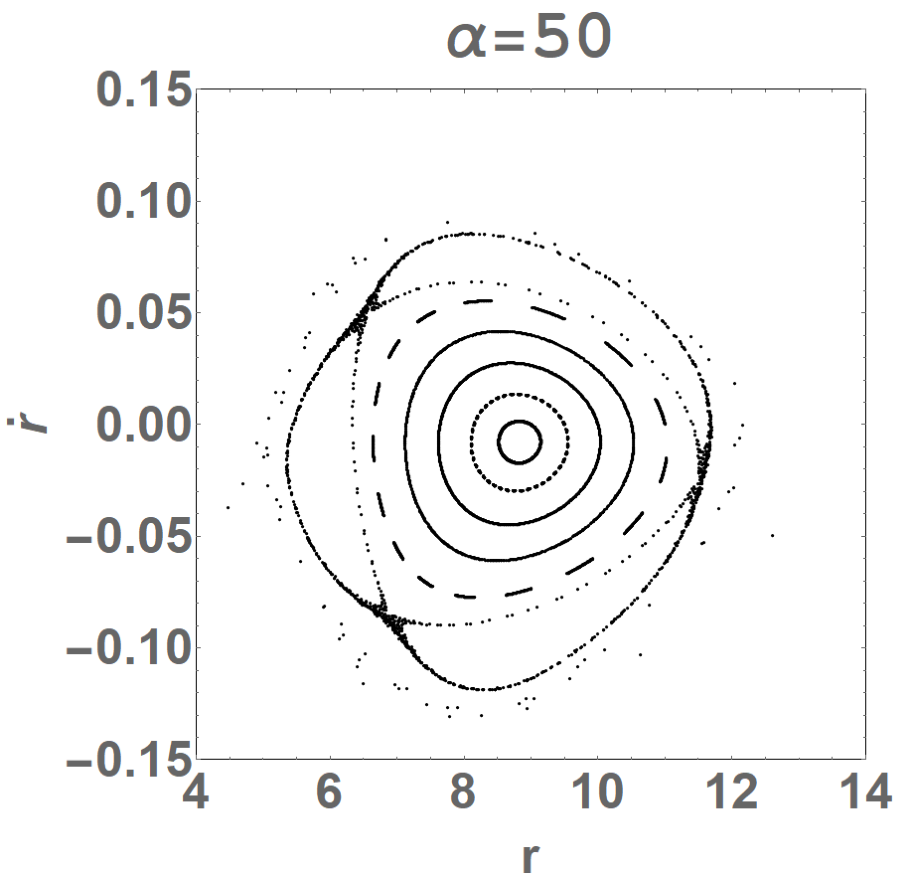}
\includegraphics[width=5.5cm ]{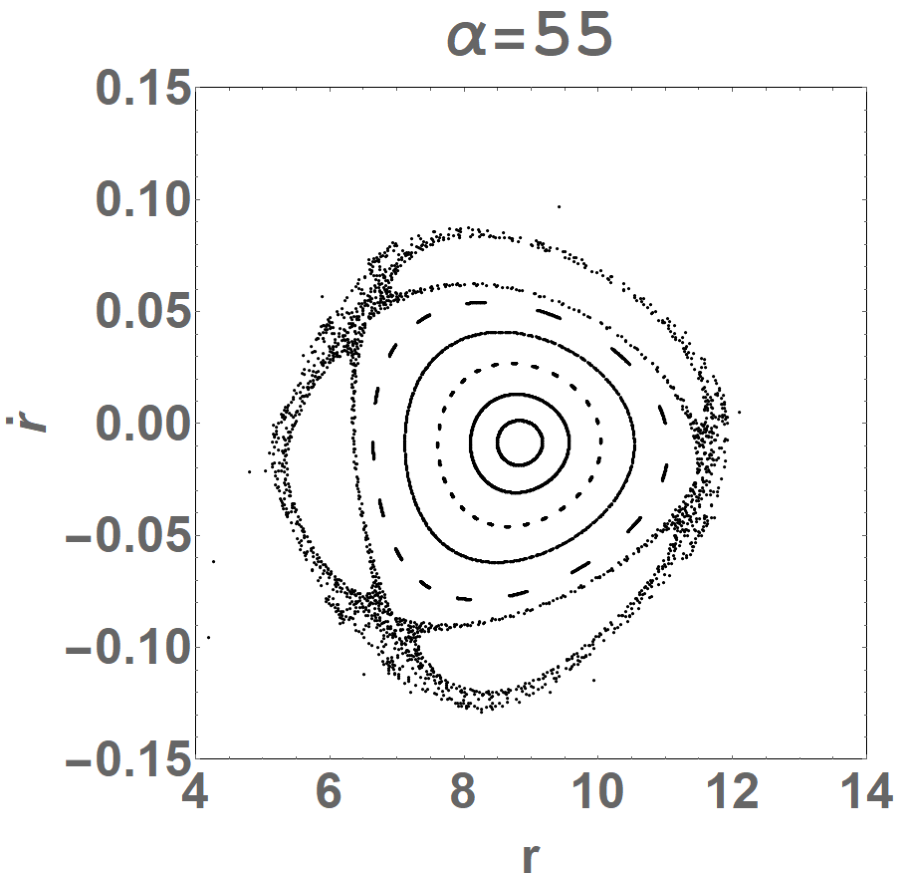}
\includegraphics[width=5.5cm ]{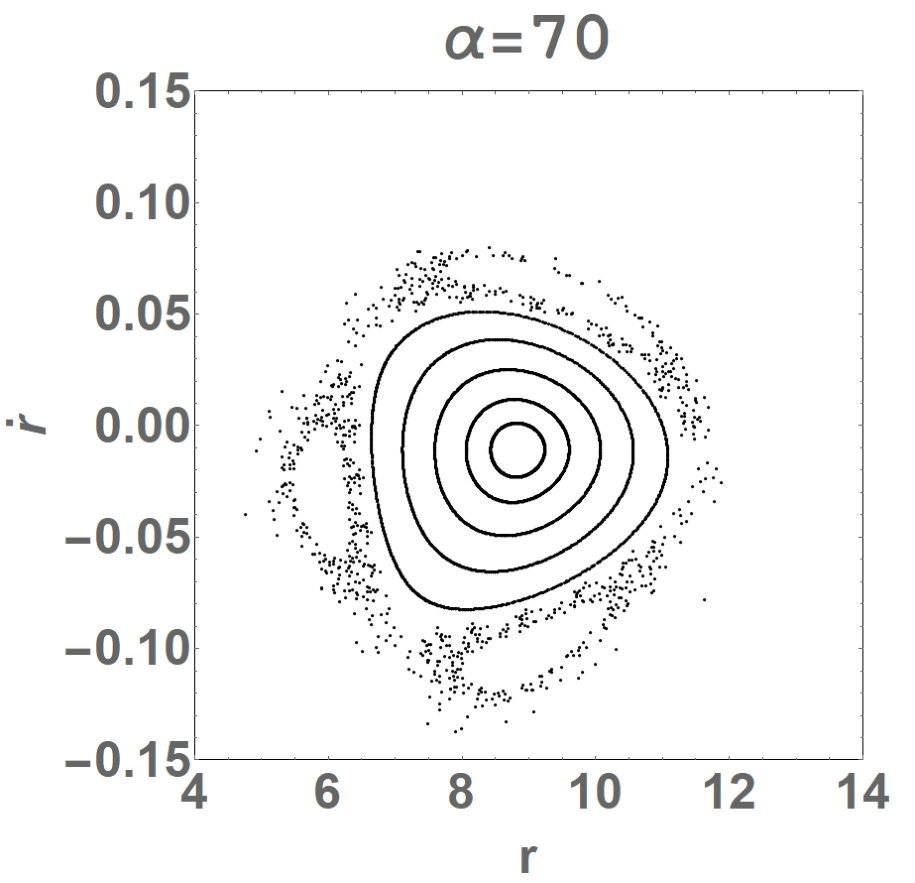}
\includegraphics[width=5.5cm ]{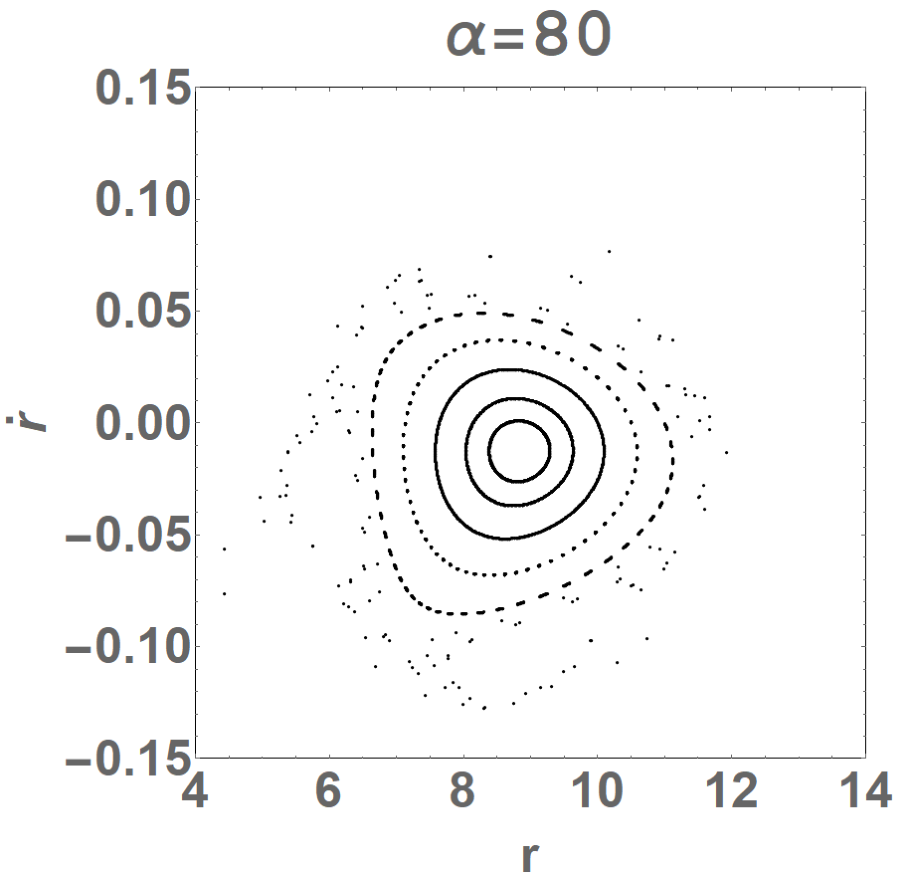}
\includegraphics[width=5.5cm ]{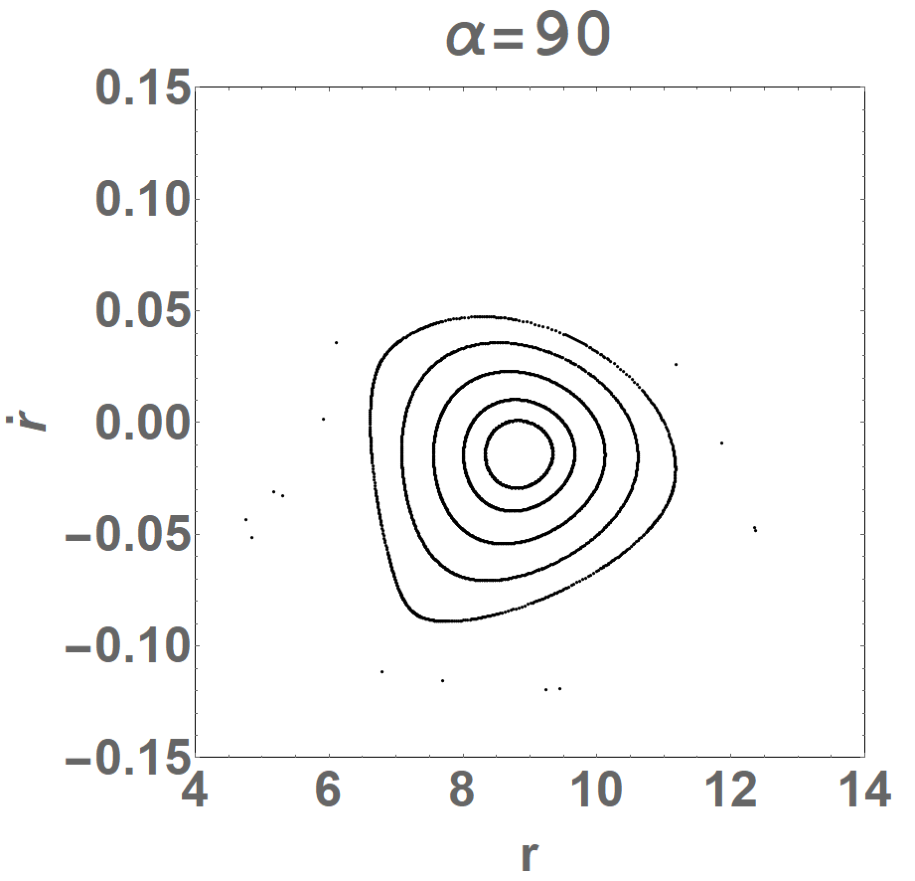}
\includegraphics[width=5.5cm ]{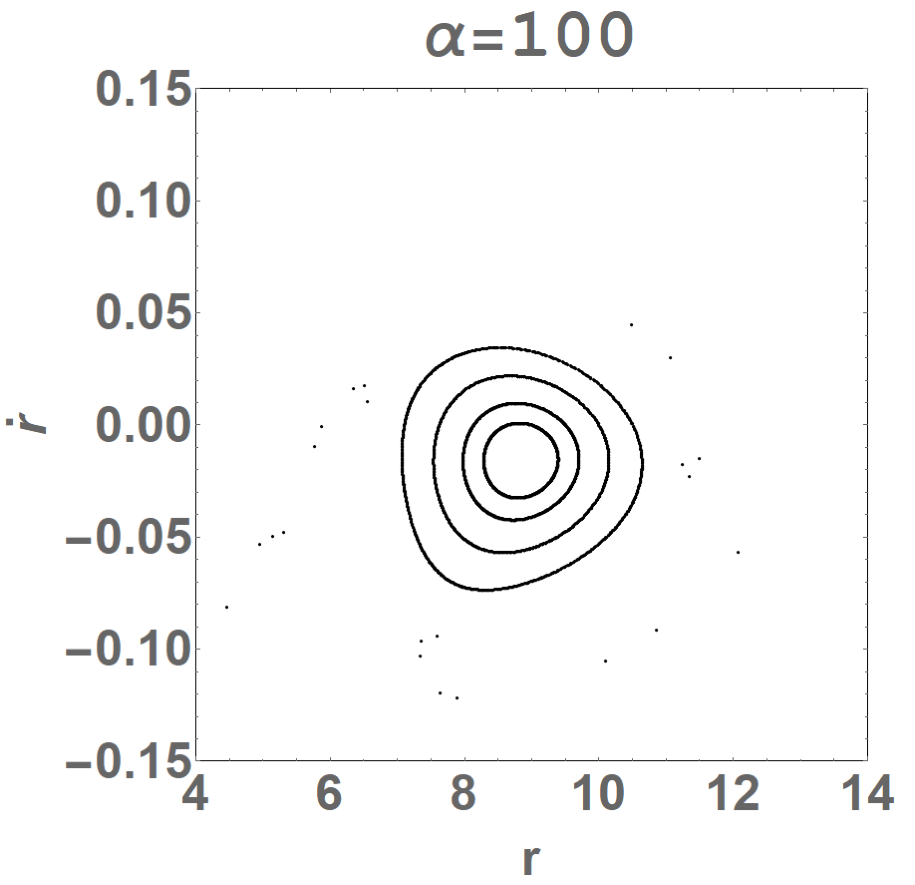}
\caption{The Poincar\'e section ($\theta=\frac{\pi}{2}$) with the coupling parameter $\alpha$ for the motion of the scalar particle coupling to the CS invariant in the stationary axisymmetric EMDA black hole for the fixed parameters $a=0.31$, $D=-0.23$, $M_{ADM}=1$, $E=0.95$ and $L=3.05M$. }\label{fig2}
\end{figure}

\begin{figure}[htbp!]
\includegraphics[width=5.5cm ]{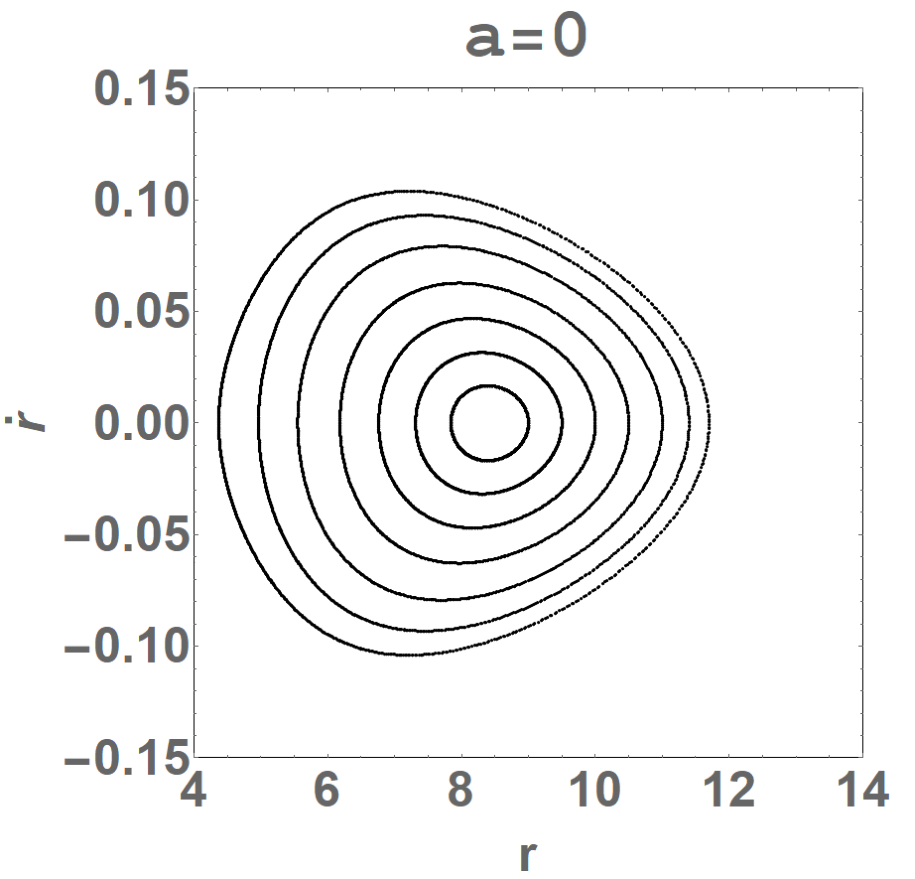}
\includegraphics[width=5.5cm ]{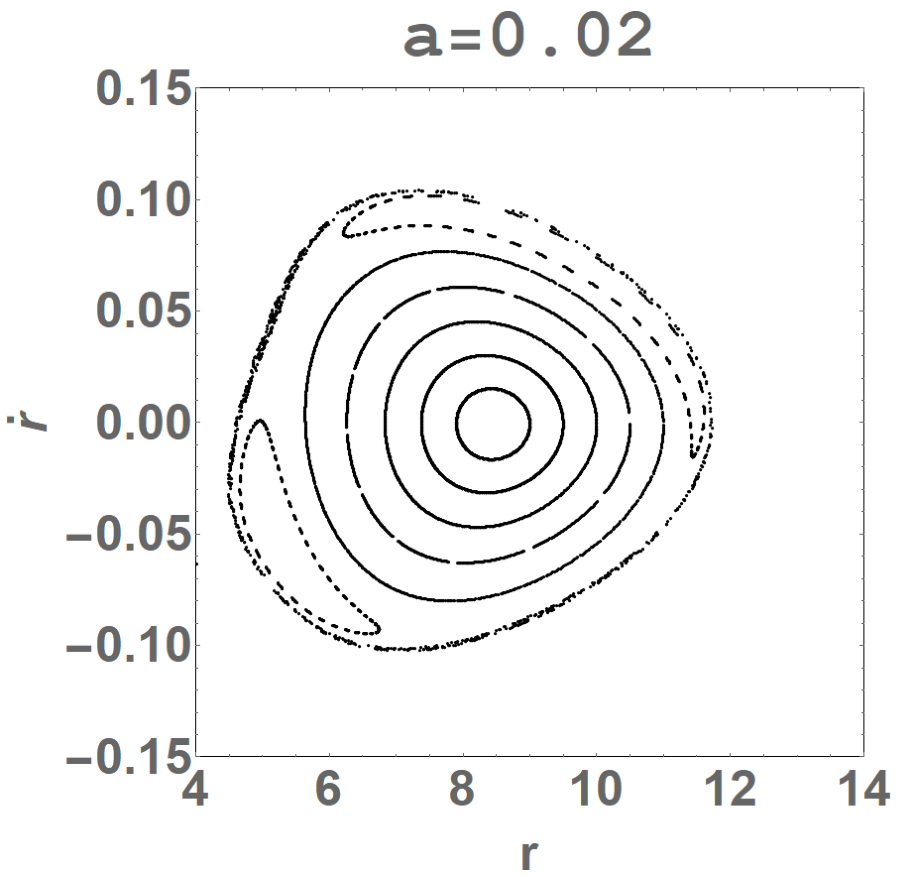}
\includegraphics[width=5.5cm ]{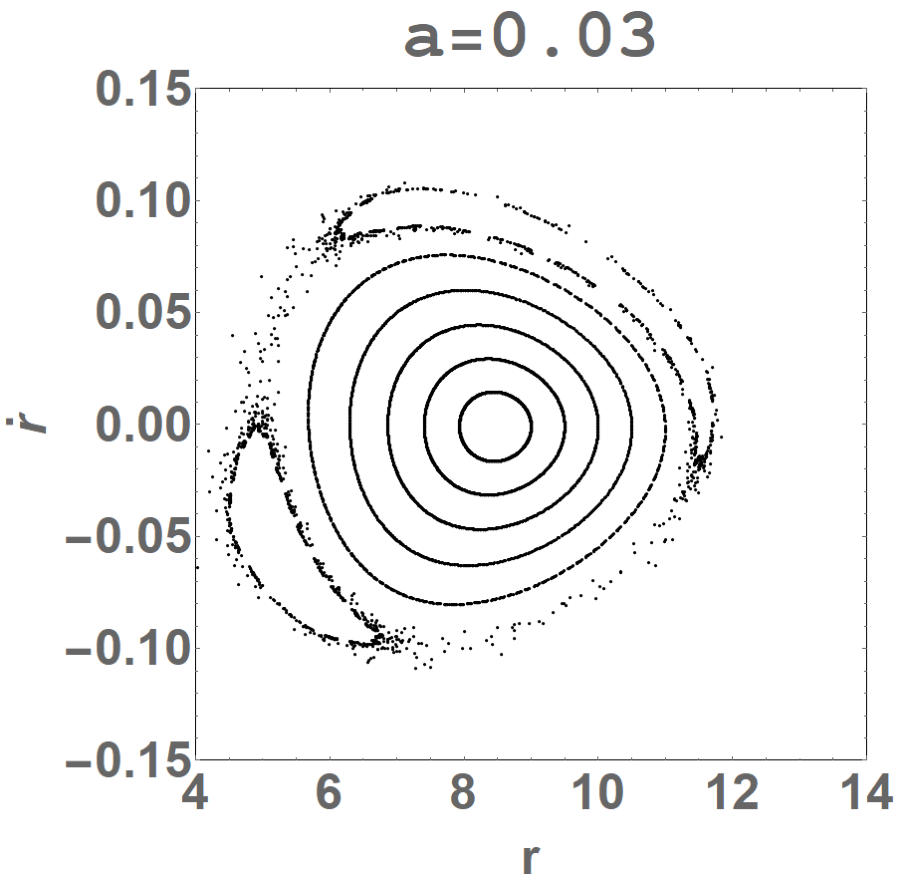}
\includegraphics[width=5.5cm ]{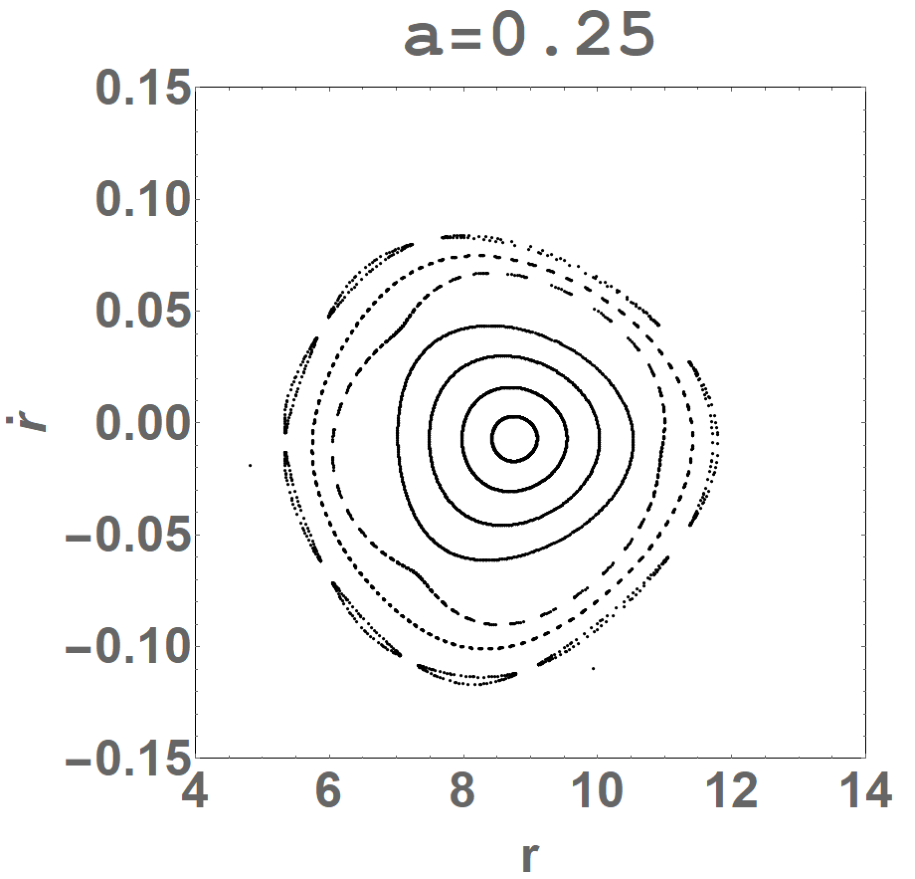}
\includegraphics[width=5.5cm ]{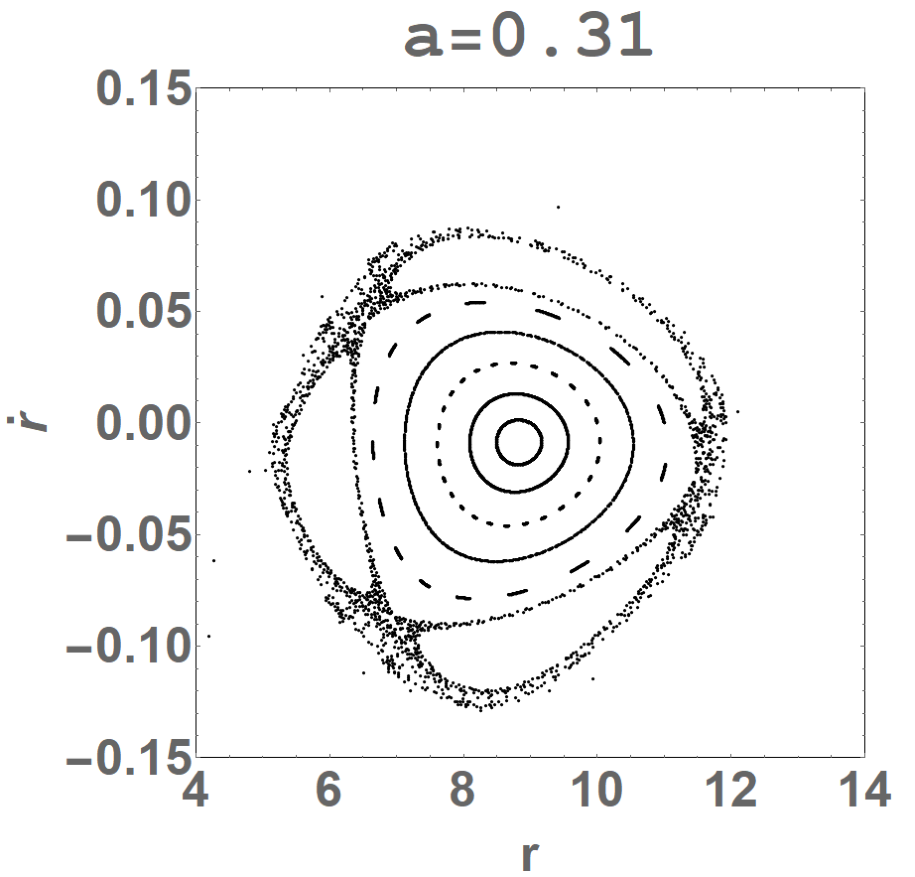}
\includegraphics[width=5.5cm ]{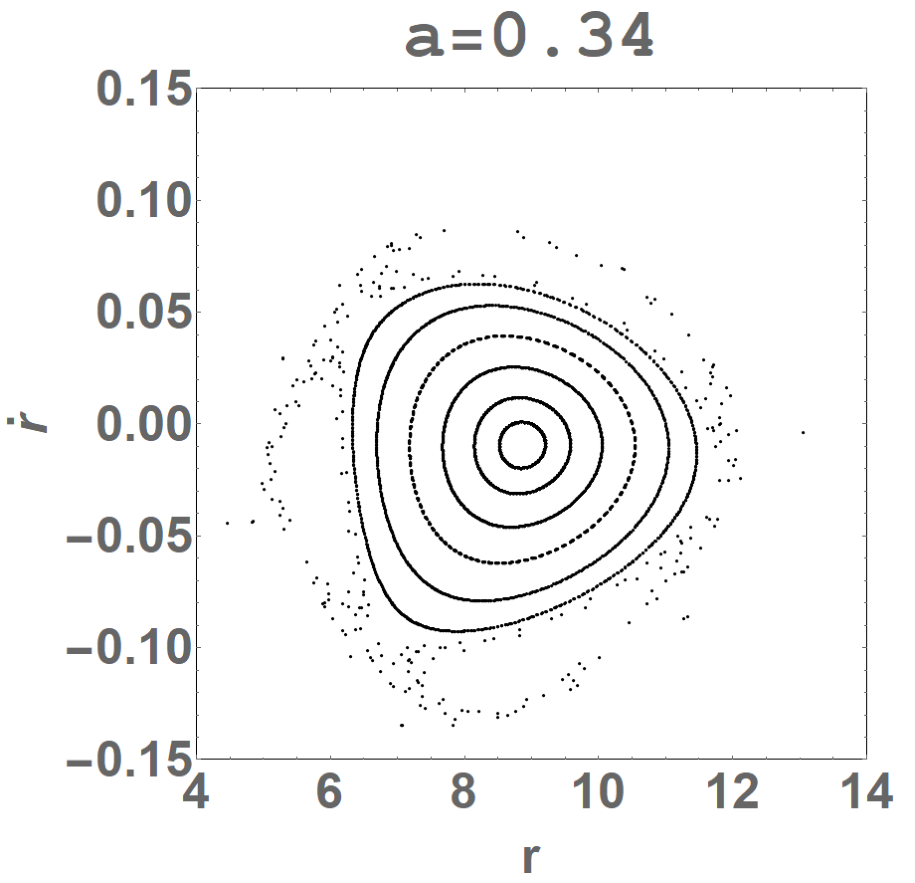}
\includegraphics[width=5.5cm ]{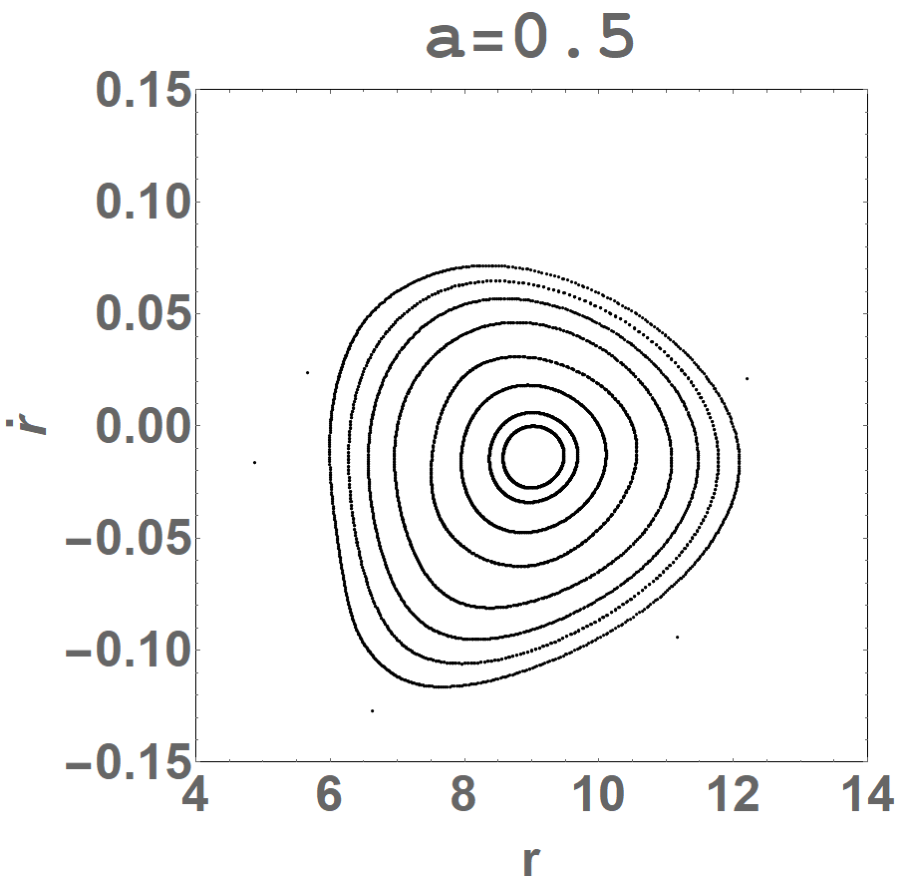}
\includegraphics[width=5.5cm ]{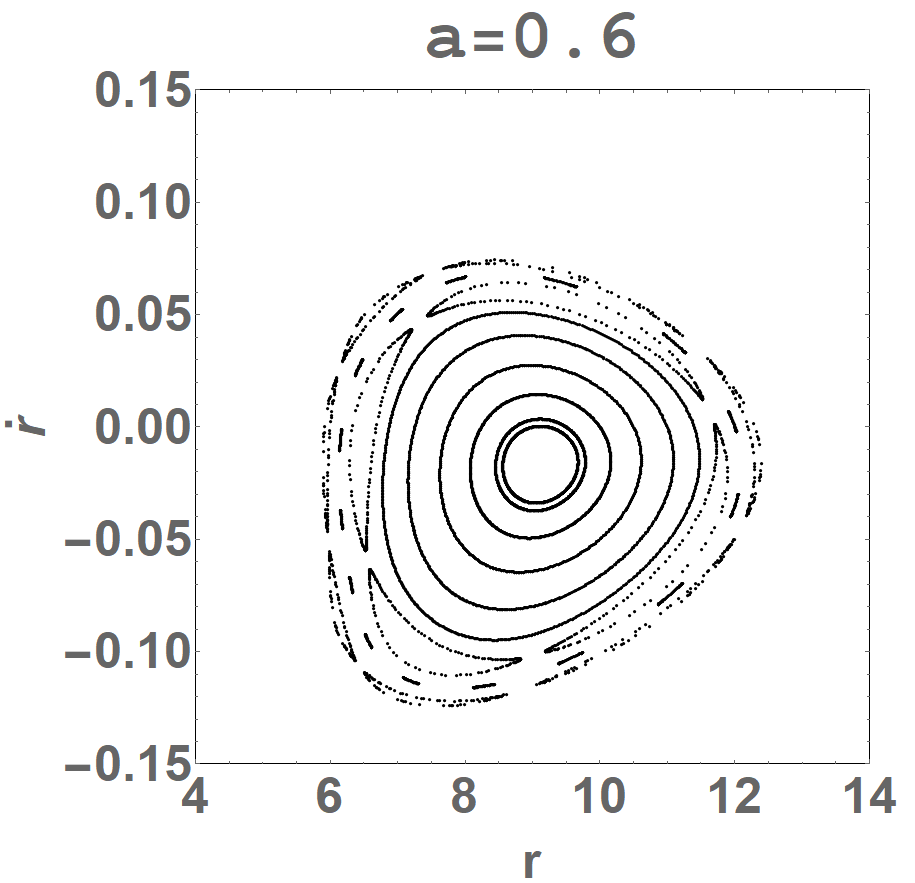}
\includegraphics[width=5.5cm ]{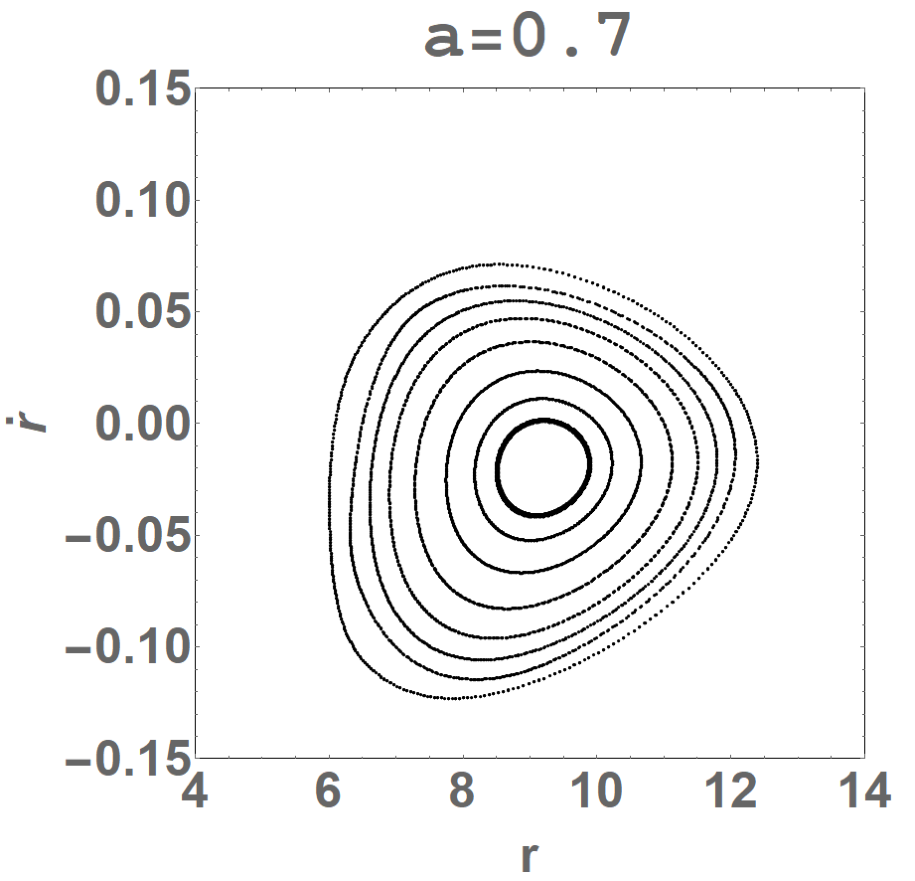}
\caption{The Poincar\'e section ($\theta=\frac{\pi}{2}$) with the spin parameter $a$ for the motion of the scalar particle coupling to the CS invariant in the stationary axisymmetric EMDA black hole for the fixed parameters $ \alpha=55$, $D=-0.23$, $M_{ADM}=1$, $E=0.95$ and $L=3.05M$. }\label{fig3}
\end{figure}

\begin{figure}[htbp!]
\includegraphics[width=5.5cm ]{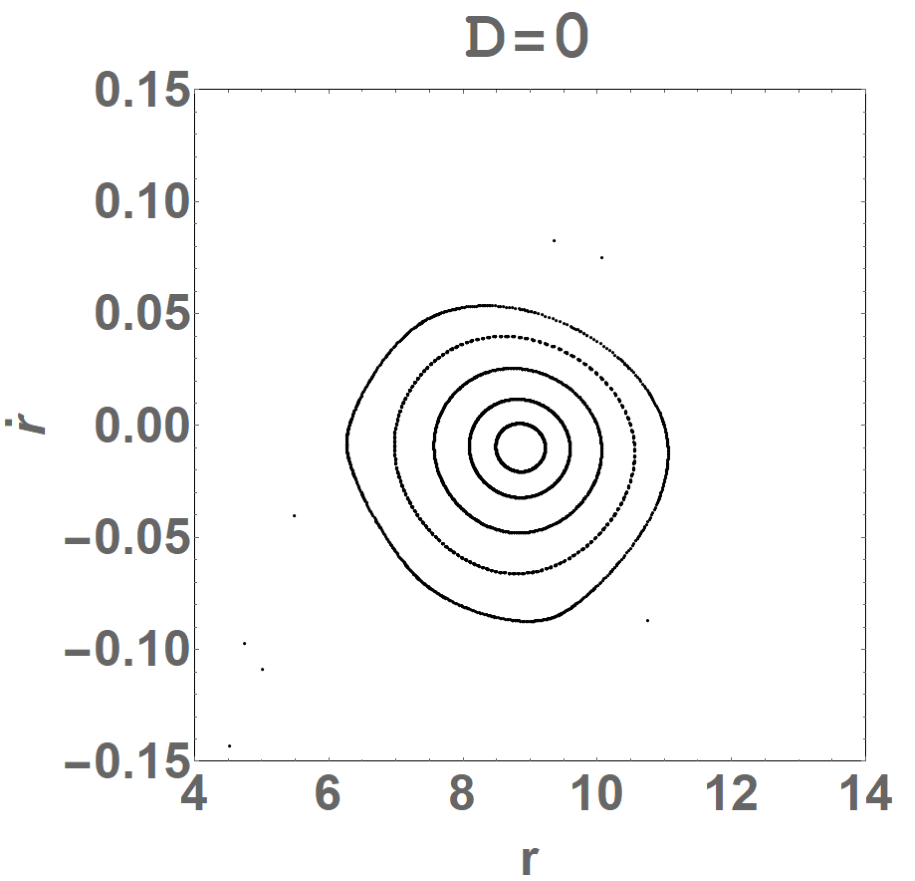}
\includegraphics[width=5.5cm ]{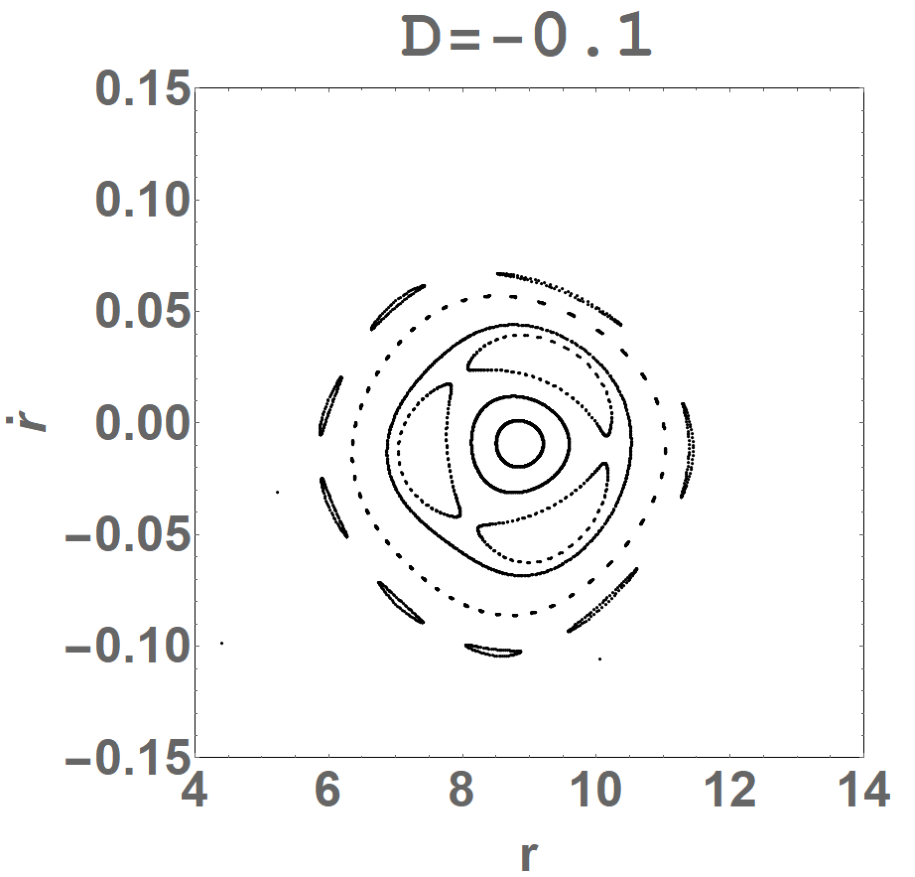}
\includegraphics[width=5.5cm ]{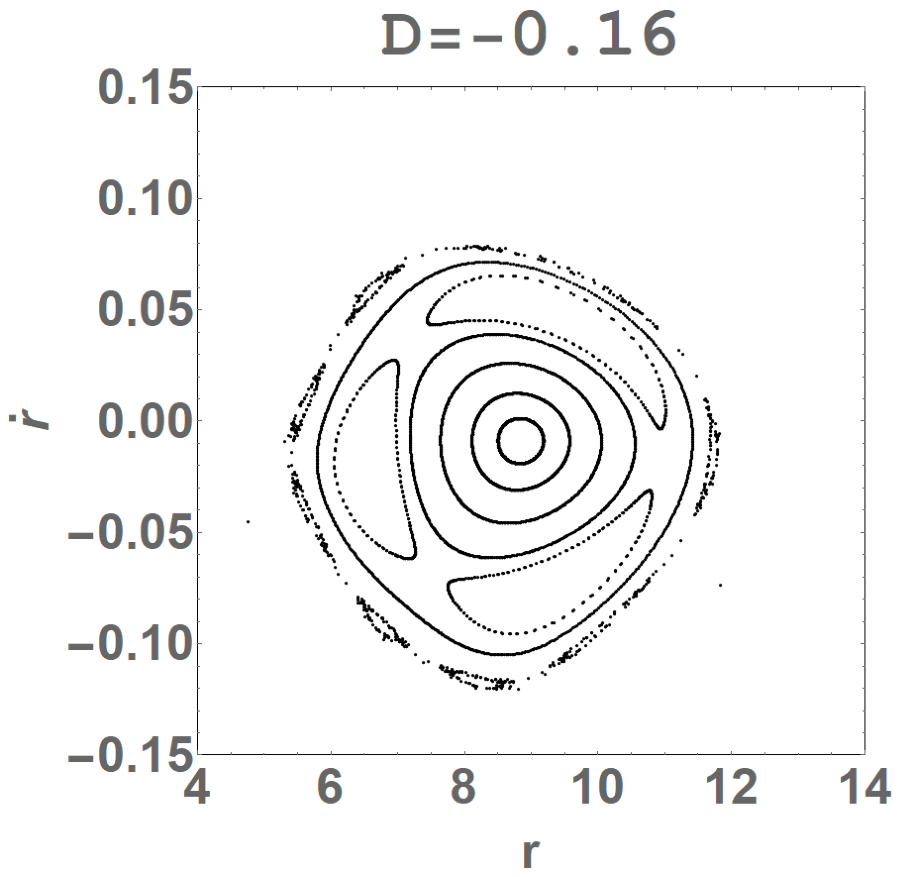}
\includegraphics[width=5.5cm ]{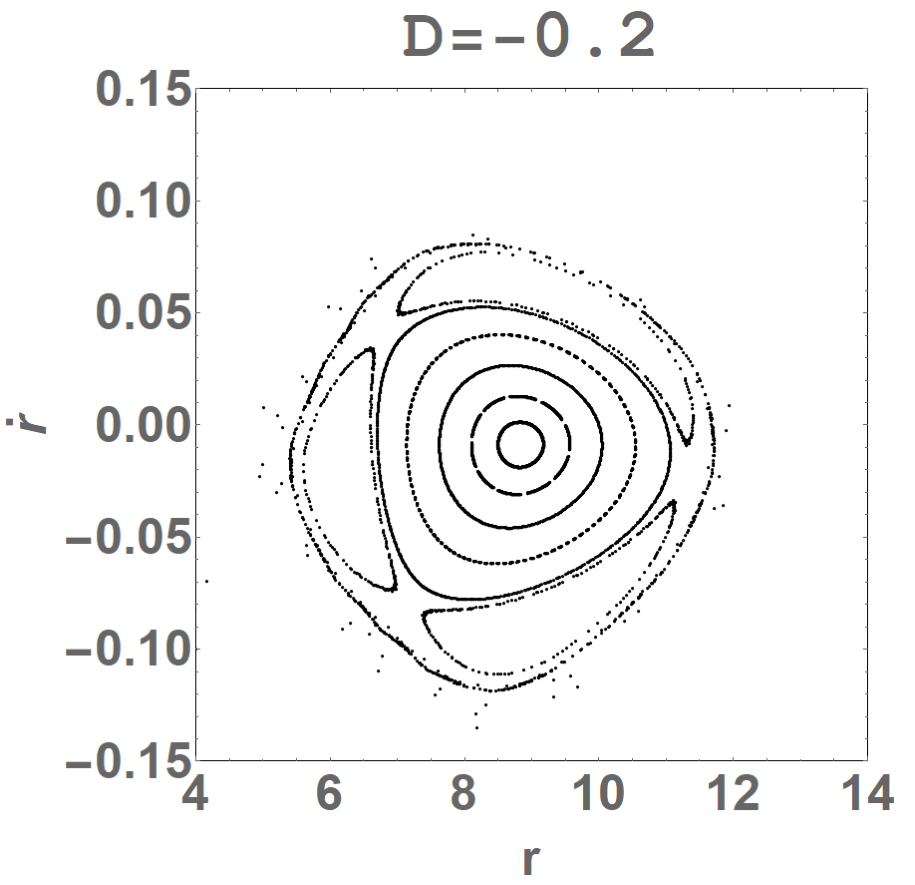}
\includegraphics[width=5.5cm ]{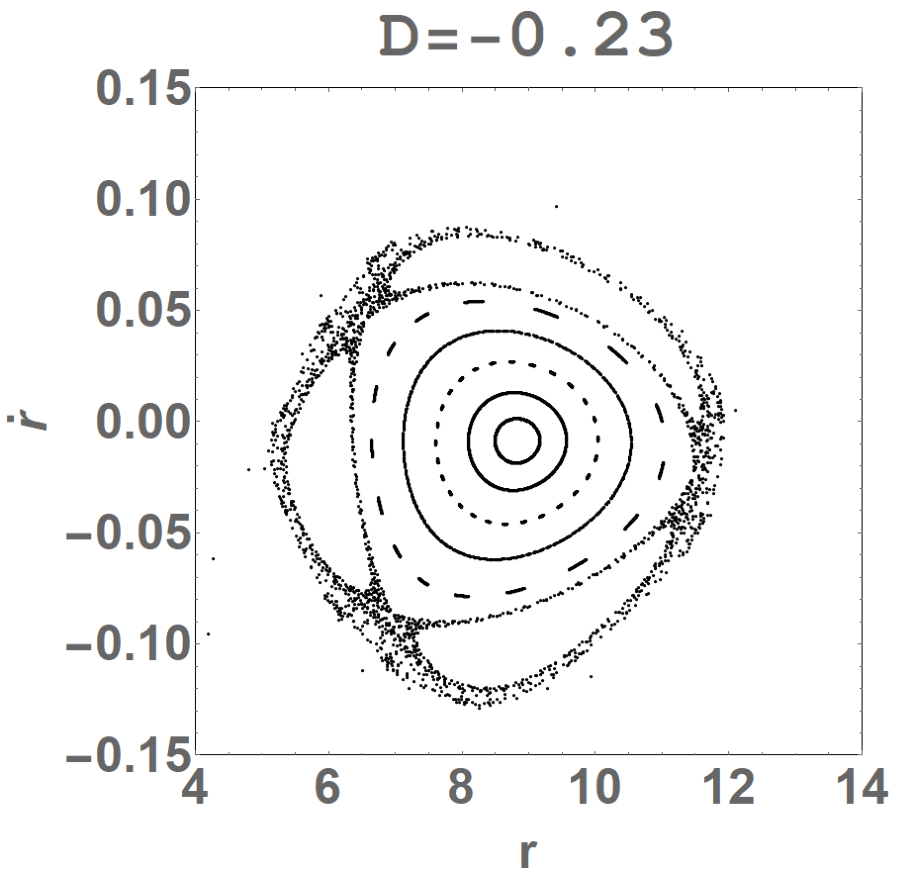}
\includegraphics[width=5.5cm ]{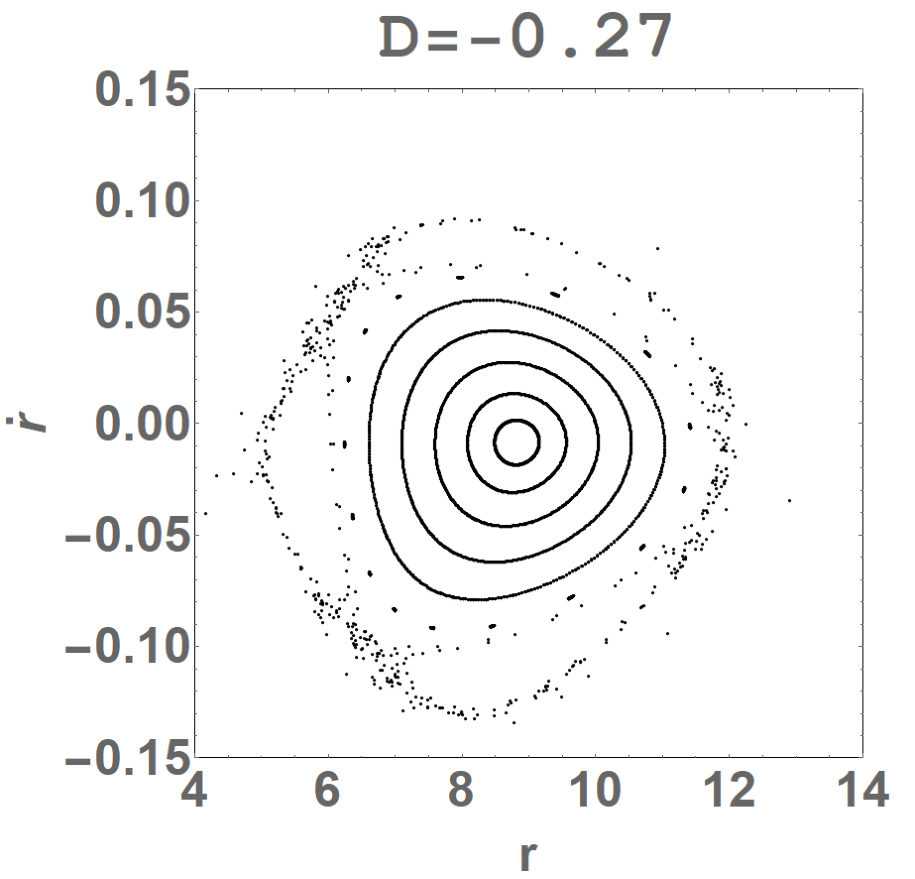}
\includegraphics[width=5.5cm ]{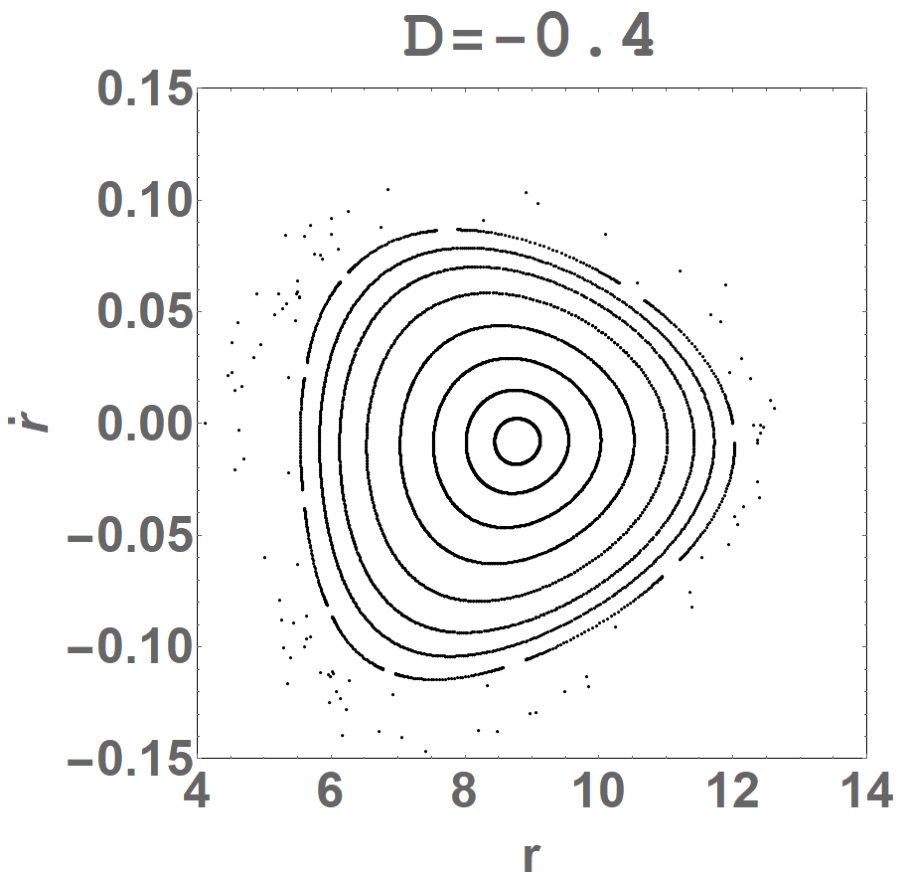}
\includegraphics[width=5.5cm ]{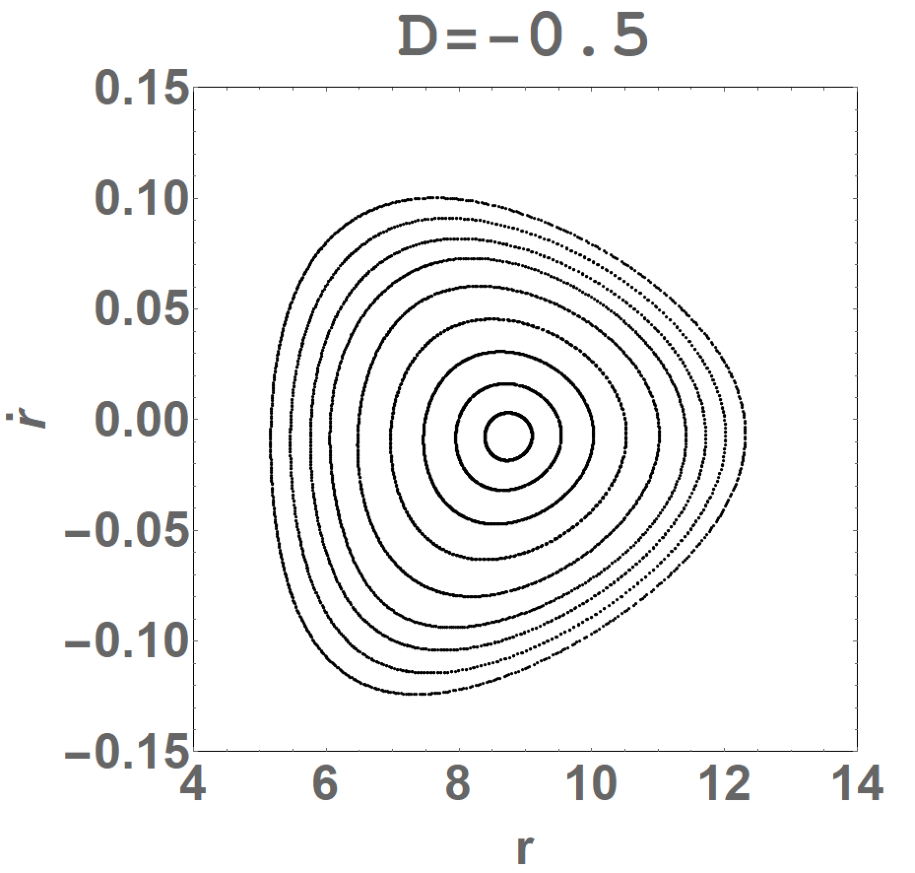}
\includegraphics[width=5.5cm ]{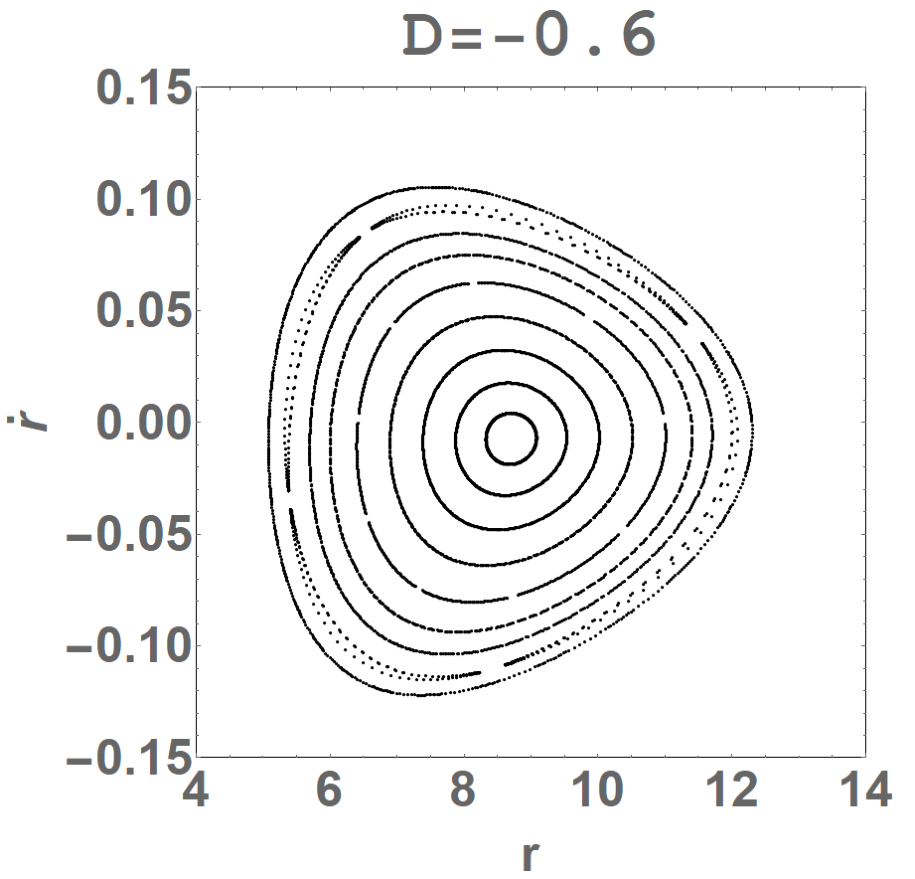}
\caption{The Poincar\'e section ($\theta=\frac{\pi}{2}$) with the dilaton parameter $D$ for the motion of the scalar particle coupling to the CS invariant in the stationary axisymmetric EMDA black hole for the fixed parameters $ \alpha=55$, $a=0.31$,  $M_{ADM}=1$, $E=0.95$ and $L=3.05M$. }\label{fig4}
\end{figure}

In Figs. \ref{fig2}-\ref{fig4}, we present the Poincar\'e section ($\theta = \frac{\pi}{2}$) containing
 nine motion orbits of the coupled particles in the background of a stationary axisymmetric EMDA black hole. We find that the changes of the regular orbit number and chaotic orbit number depend on the black hole parameters $a$, $D$ and the coupling parameter $\alpha$. For the fixed spin parameter $a=0.31$ and dilaton parameter $D=-0.23$, we find that all of nine orbits are regular as $\alpha=0$ because in this case the motion equations of scalar test particles reduce to
the usual variable-separable geodesic equations. With the increase of  $\alpha$, the number of regular motion orbits shrinks and the number of chaotic orbits increases, which is similar to that in the Kerr black hole case \cite{Zhou}. Moreover, we also find that the chaotic orbits are farther from the central fixed point than the regular orbits.
For the fixed $\alpha=55$ and $D=-0.23$, when $a=0$, one can find that there exist only the regular orbits and the chaos does not occur as in the case $\alpha=0$, which is because that the CS invariant $^* RR$ contains a factor of $a$ and it disappears for the static black hole. With the increase of the spin parameter $a$ of the black hole, we find that the number of regular orbits first decreases and then increases. For the fixed $\alpha=55$ and $a=0.31$, increasing the absolute value of dilaton parameter $D$, we observe that the number of regular orbits first increases and then decreases, and finally increases again. Meanwhile, the number of chaotic orbits first increases and then decreases. Moreover, with the increasing $\alpha$, $a$ and $|D|$, we also find the chaotic strength for the chaotic orbits first increases and then decreases. Thus, the coupling together with the spin and dilaton parameters yields the richer dynamical behavior of the scalar particle in the stationary axisymmetric EMDA black hole spacetime. As in the Kerr case \cite{Zhou}, we also note from Figs. \ref{fig1}-\ref{fig4} that the patterns in
the Poincar\'{e} section lose the reflection symmetry along the axial line $\dot{r}=0$
due to the interaction with the CS invariant, which could be a common feature for the particle motions under such a coupling. This can be attributed to that the CS invariant $^* RR$ (\ref{CSRR1}) is not symmetric with respect to the equatorial plane.


\begin{figure}[ht]
\includegraphics[width=5.5cm ]{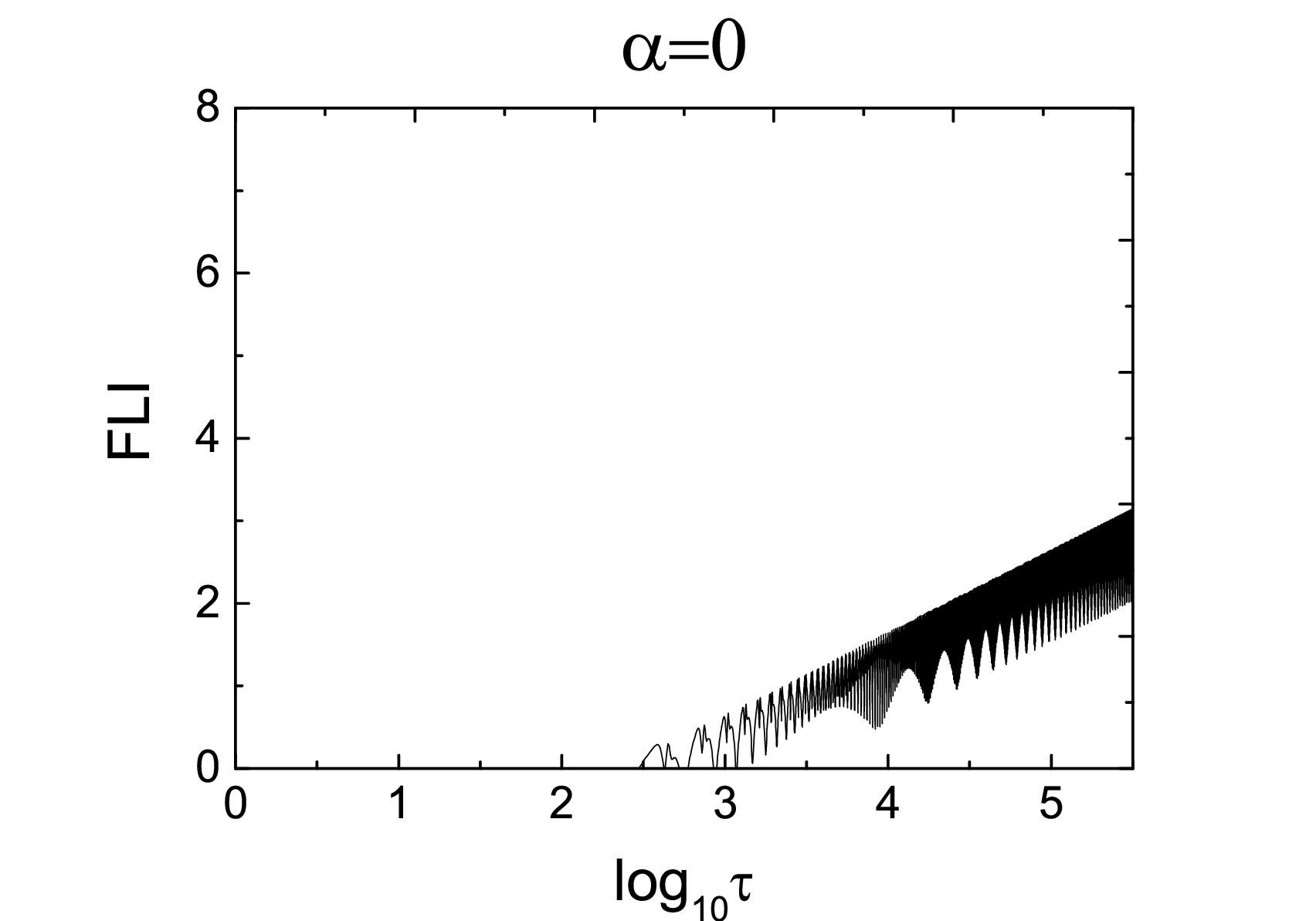}
\includegraphics[width=5.5cm ]{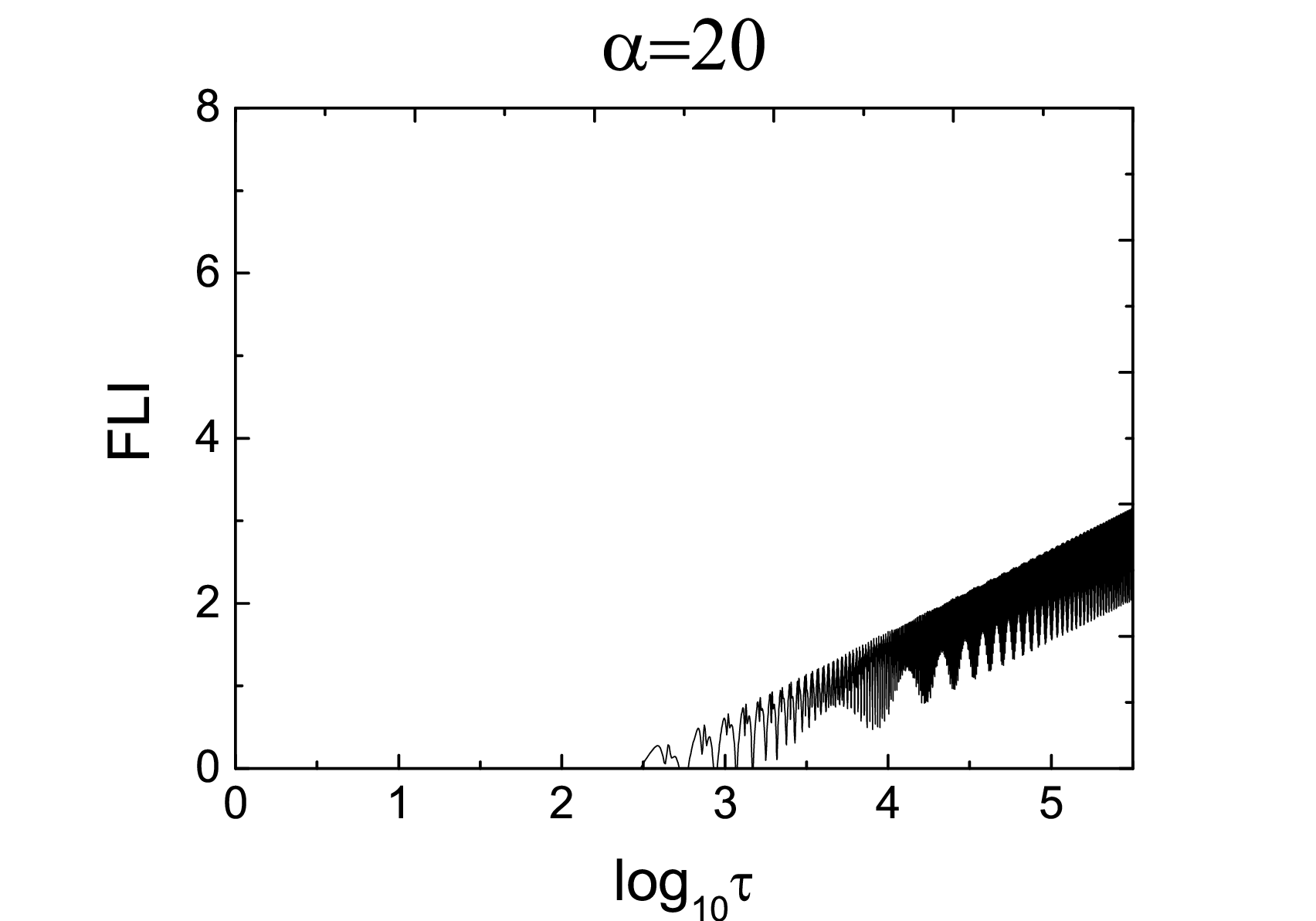}
\includegraphics[width=5.5cm ]{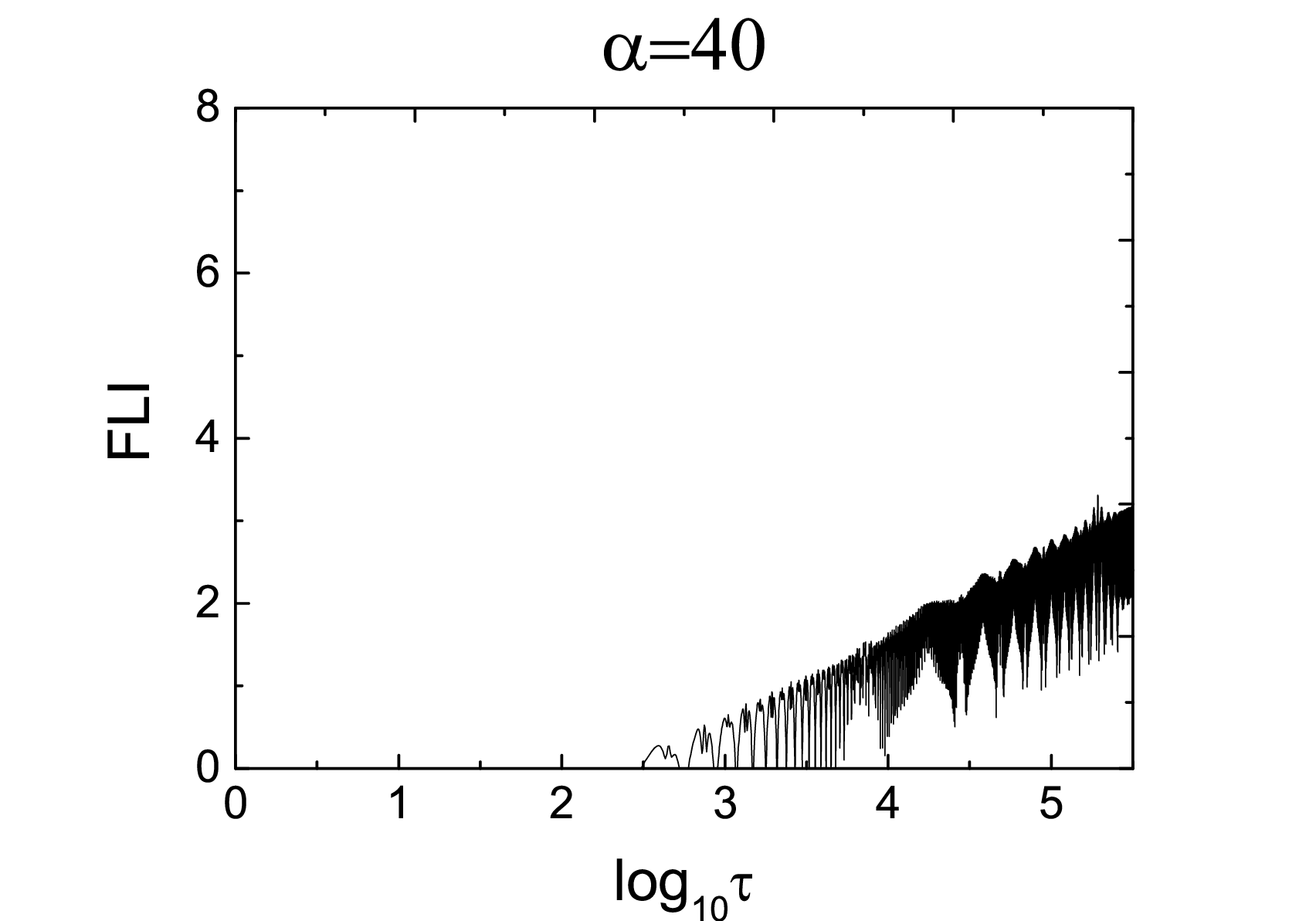}
\includegraphics[width=5.5cm ]{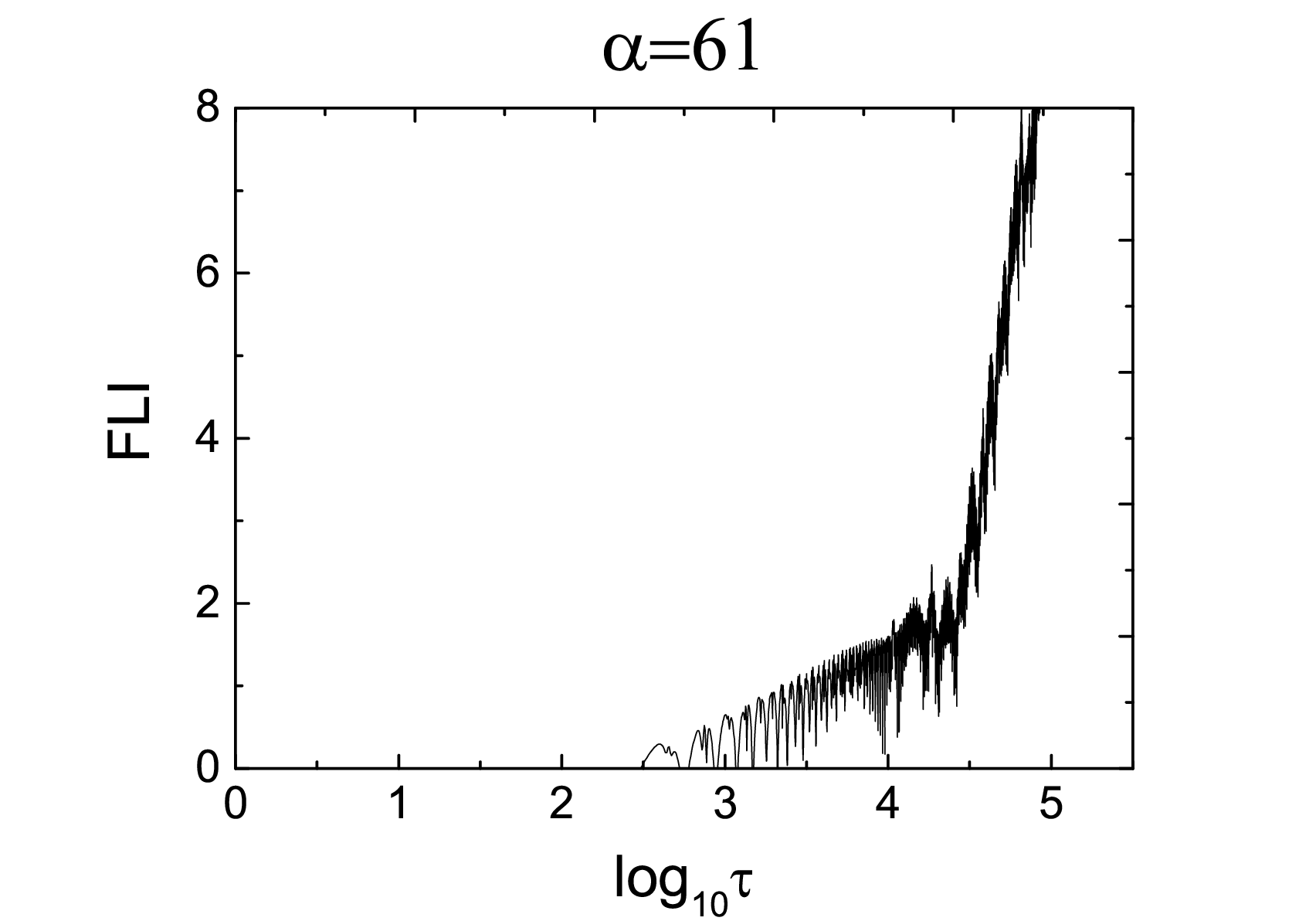}
\includegraphics[width=5.5cm ]{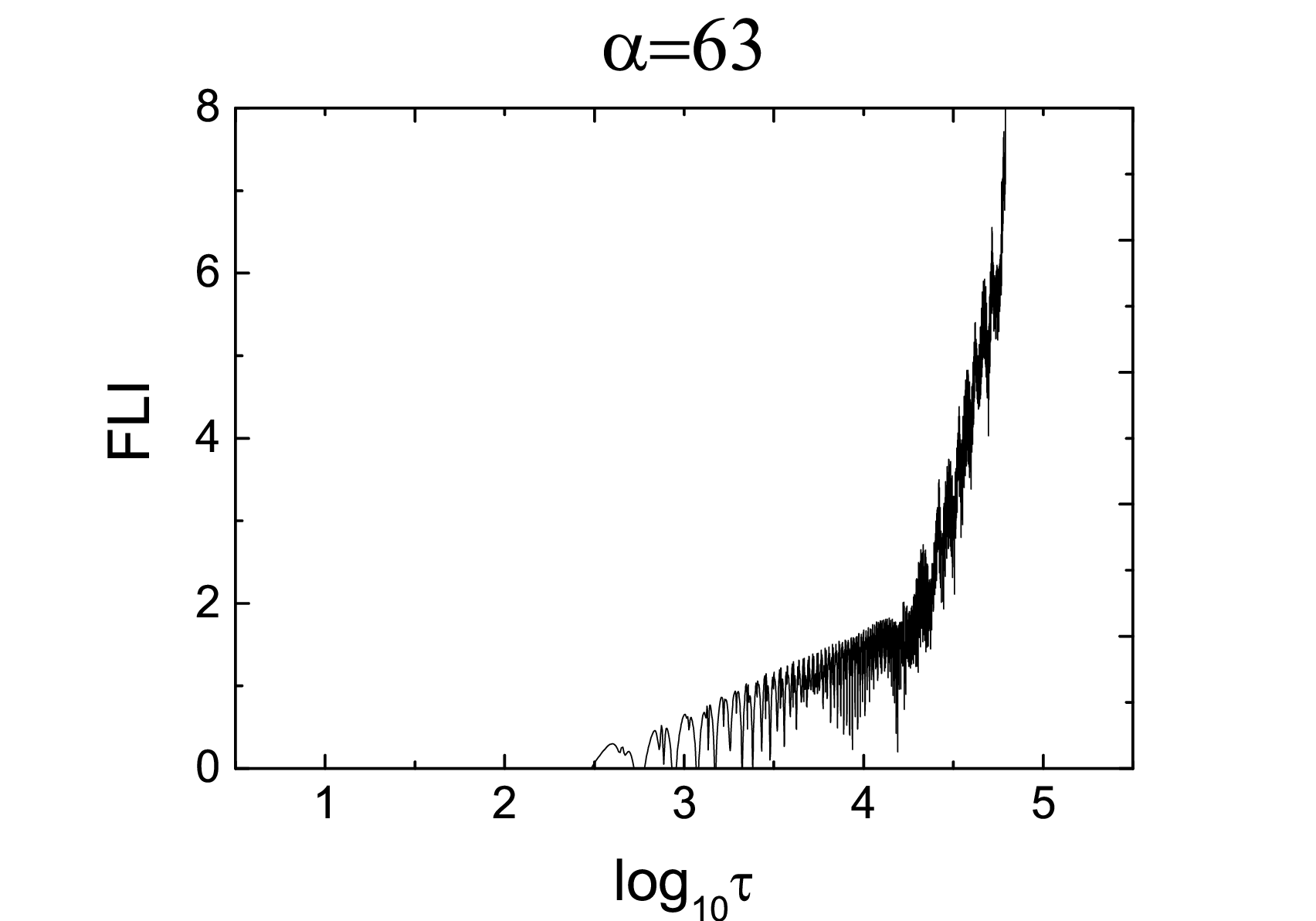}
\includegraphics[width=5.5cm ]{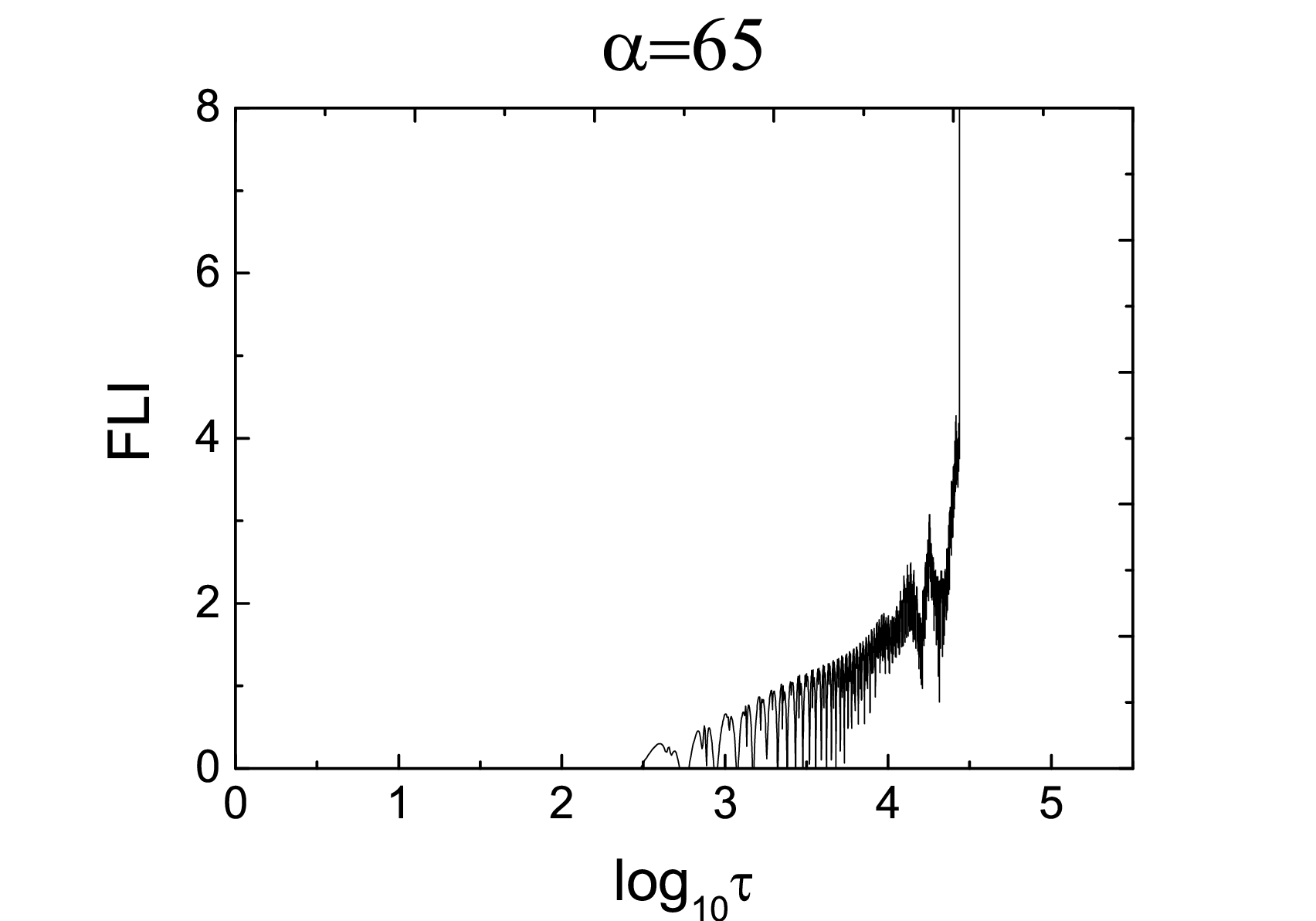}
\caption{The fast Lyapunov Indicator (FLI) with the coupling parameter $\alpha$ for the signals shown in Fig. \ref{fig1}. }\label{fig5}
\end{figure}

FLI is a fast and efficient tool to identify the chaotic behavior of the particle. In a curved spacetime, the FLI with the two-particle method can be expressed by \cite{Tancredi,Froe,Wu,Chen}
\begin{eqnarray}
FLI(\tau)=-(k+1)\ast\log_{10}d(0)+\log_{10}d(\tau),
\end{eqnarray}
where $d(\tau)=\sqrt{|g_{\mu\nu}\Delta x^{\mu}\Delta x^{\nu}}|$, $\Delta x^{\mu}$ is the deviation vector between two adjacent trajectories. To avoid numerical saturation caused by the rapid separation of two adjacent trajectories, the sequential number of renormalization $k$ is introduced. Whenever $d(\tau)=1$, the value of $k$ is increased by one and then $d(\tau)$ is pulled back to a distance of $d(0)$. The FLI($\tau$) grows exponentially for chaotic orbits, but it grows algebraically with time for the regular orbits. In Fig. \ref{fig5}, we present the variation of $\text{FLI}(\tau)$ with the coupling parameter $\alpha$ for the initial orbit selected in Fig. \ref{fig1}. It shows that the $\text{FLI}(\tau)$ increases linearly with $\tau$ as $\alpha<61$, which means that the motion of the scalar particle is regular in this case.  However, in the case of $\alpha\geq 61$, the $\text{FLI}(\tau)$ grows exponentially with $\tau$, and the corresponding motion is chaotic. These results agree with those obtained from the Poincar\'{e} section shown in Fig. \ref{fig1}.

\begin{figure}[htbp!]
\includegraphics[width=5.3cm ]{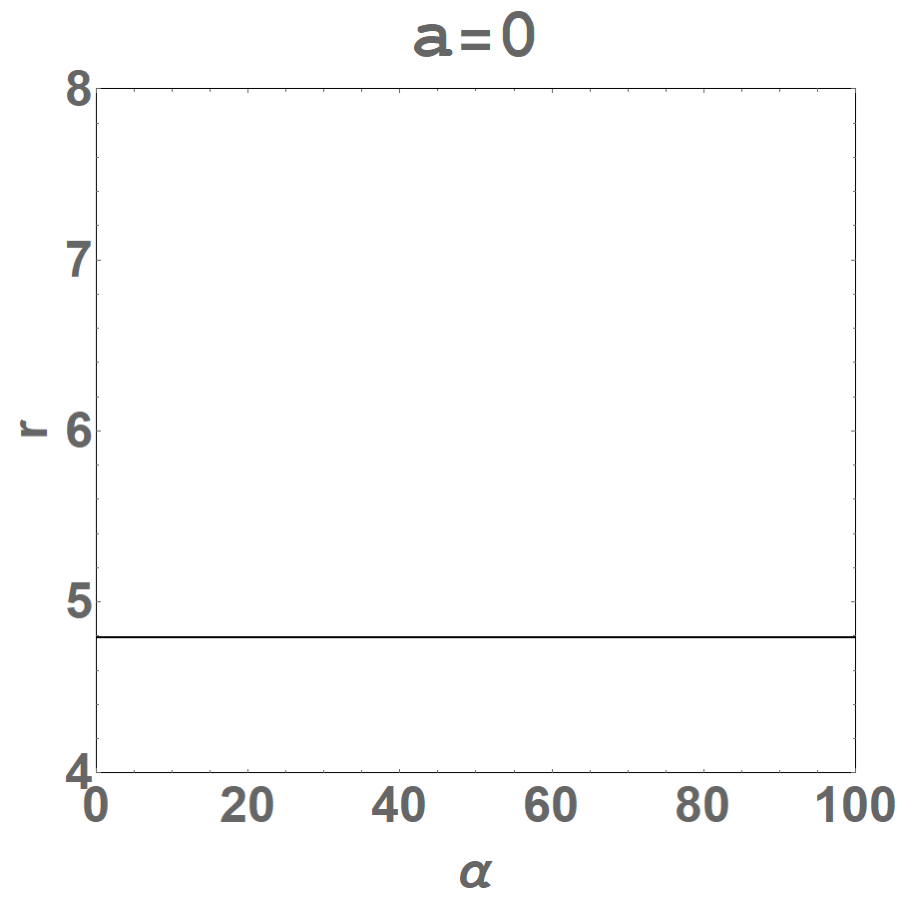}
\includegraphics[width=5.3cm ]{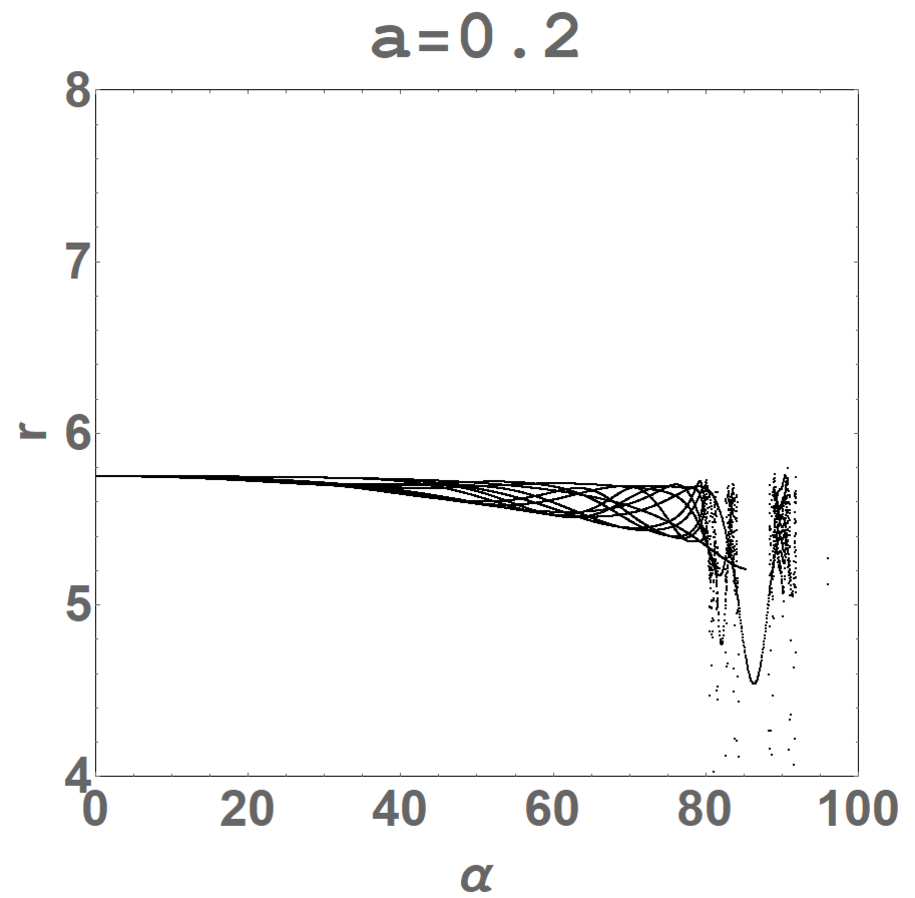}
\includegraphics[width=5.3cm ]{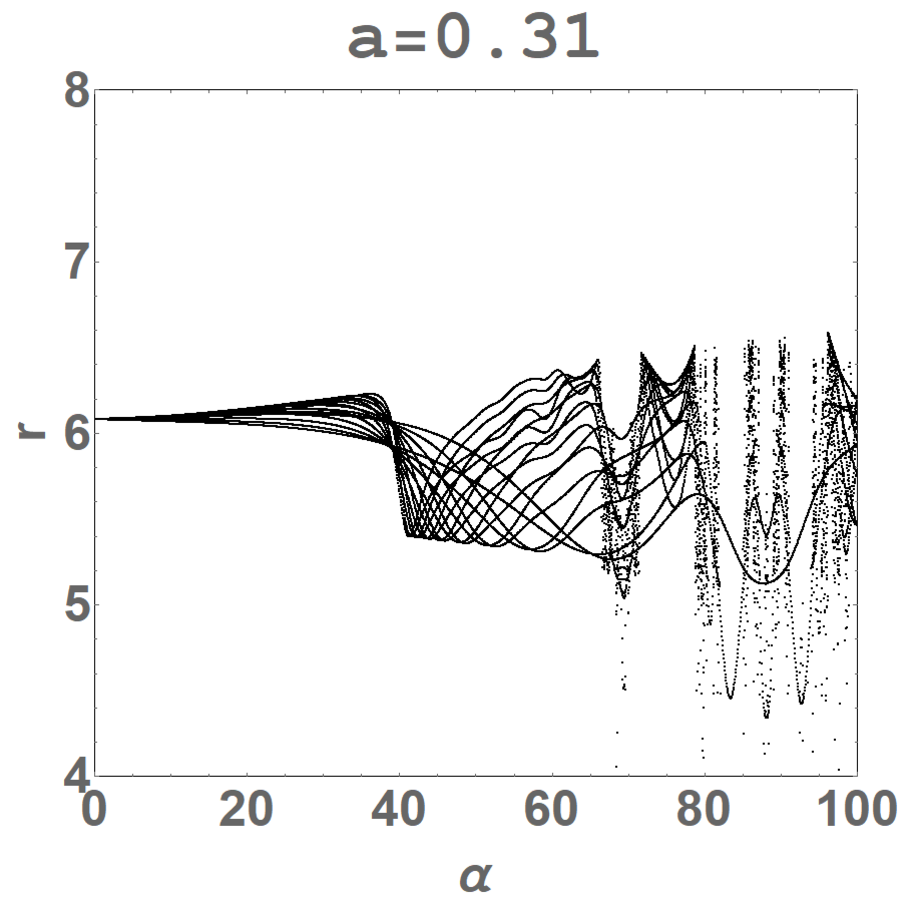}
\includegraphics[width=5.3cm ]{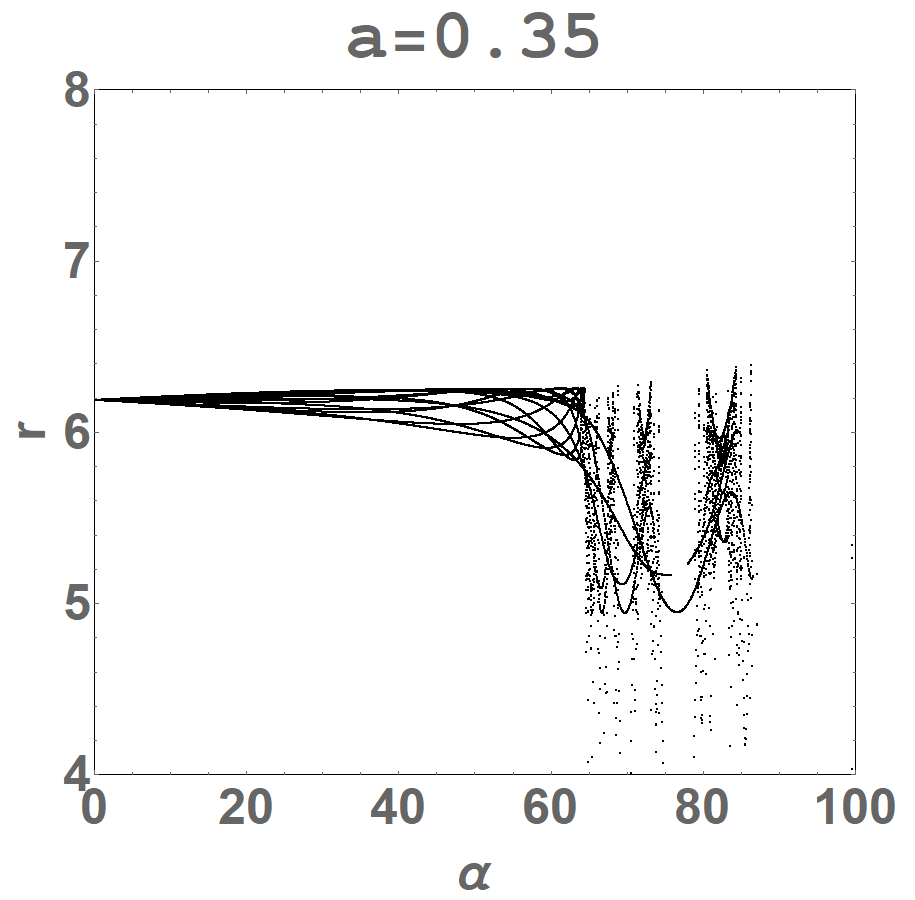}
\includegraphics[width=5.3cm ]{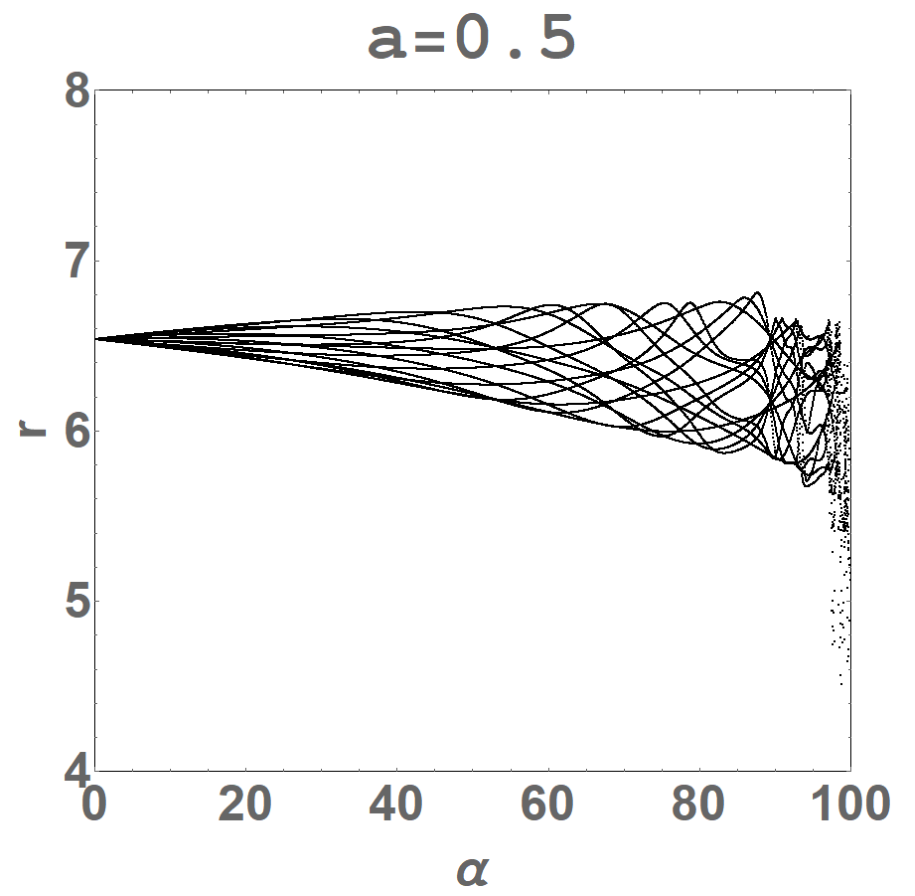}
\includegraphics[width=5.3cm ]{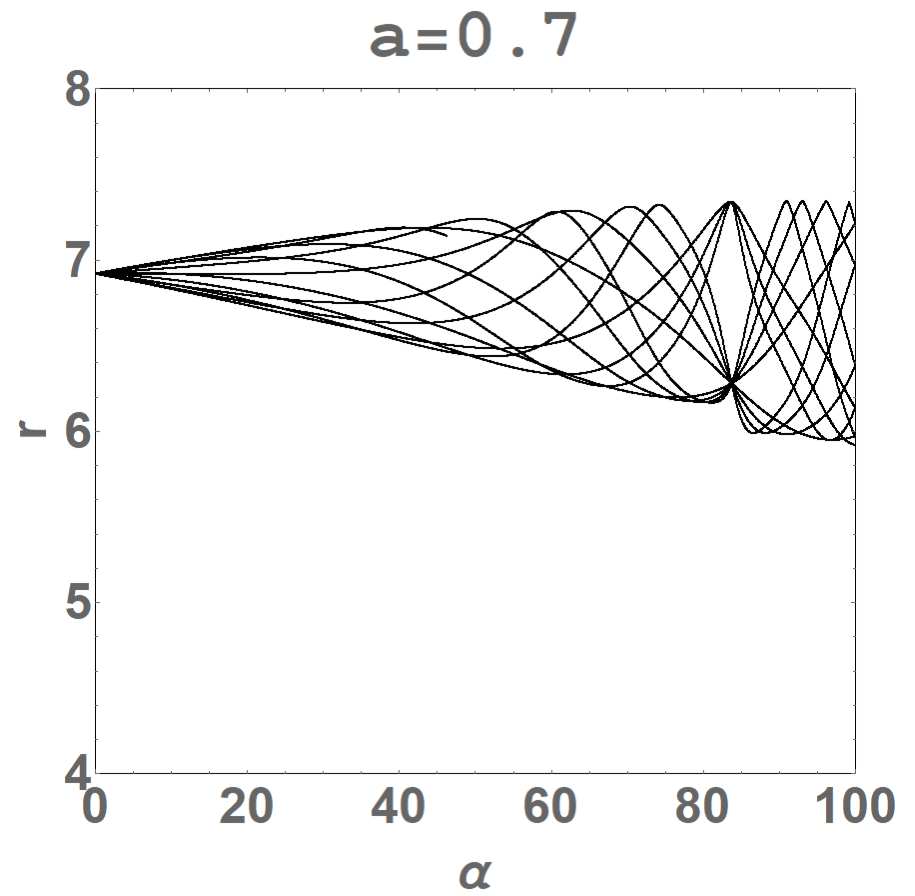}
\caption{The bifurcation changes with the CS coupling parameter $\alpha$ for the fixed dilaton parameter $D=-0.23$ and different spin parameters $a$.}\label{fig6}
\end{figure}

\begin{figure}[htbp!]
\includegraphics[width=5.3cm ]{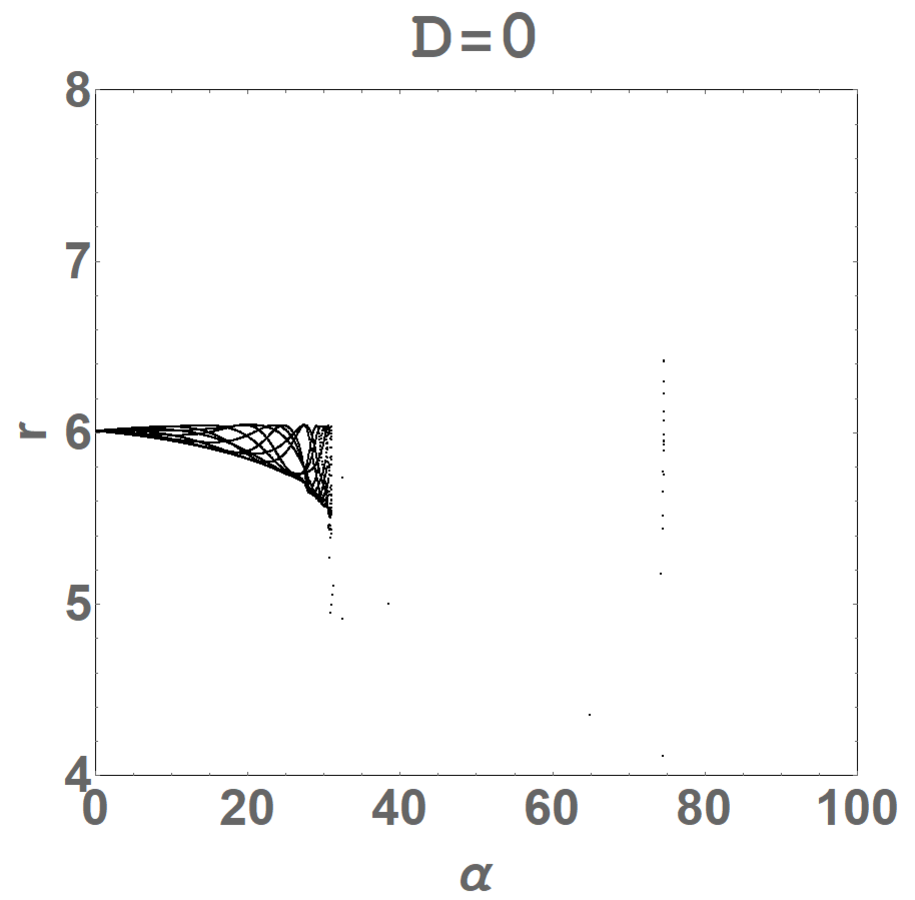}
\includegraphics[width=5.3cm ]{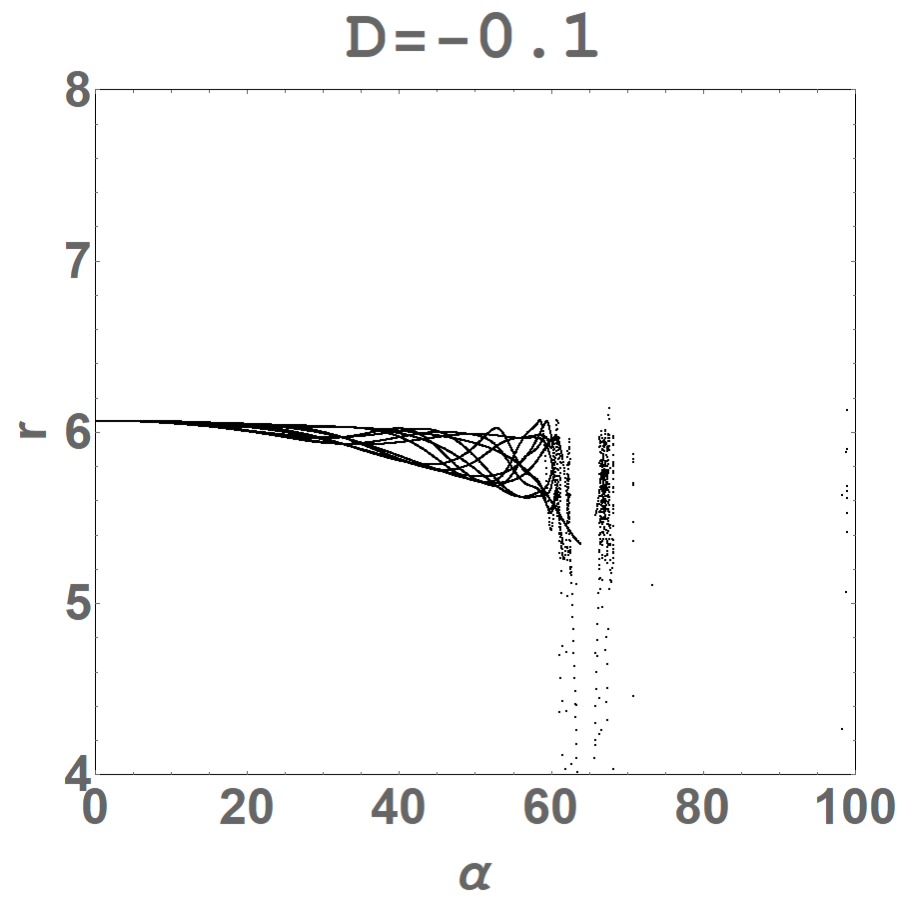}
\includegraphics[width=5.3cm ]{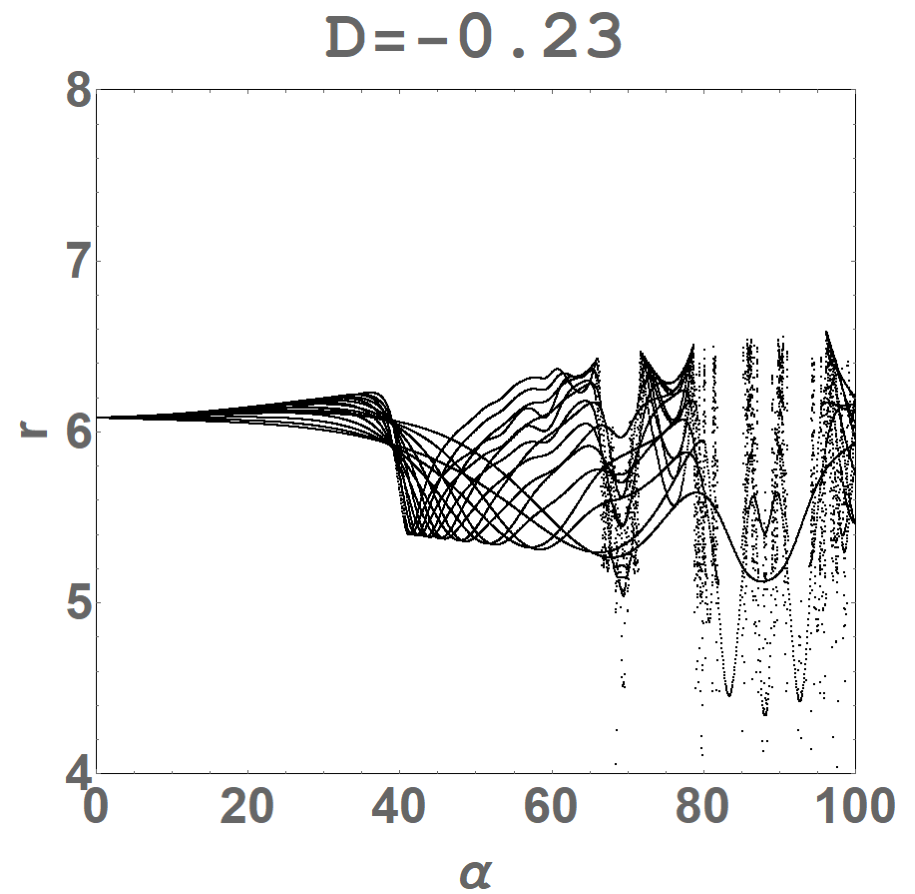}
\includegraphics[width=5.3cm ]{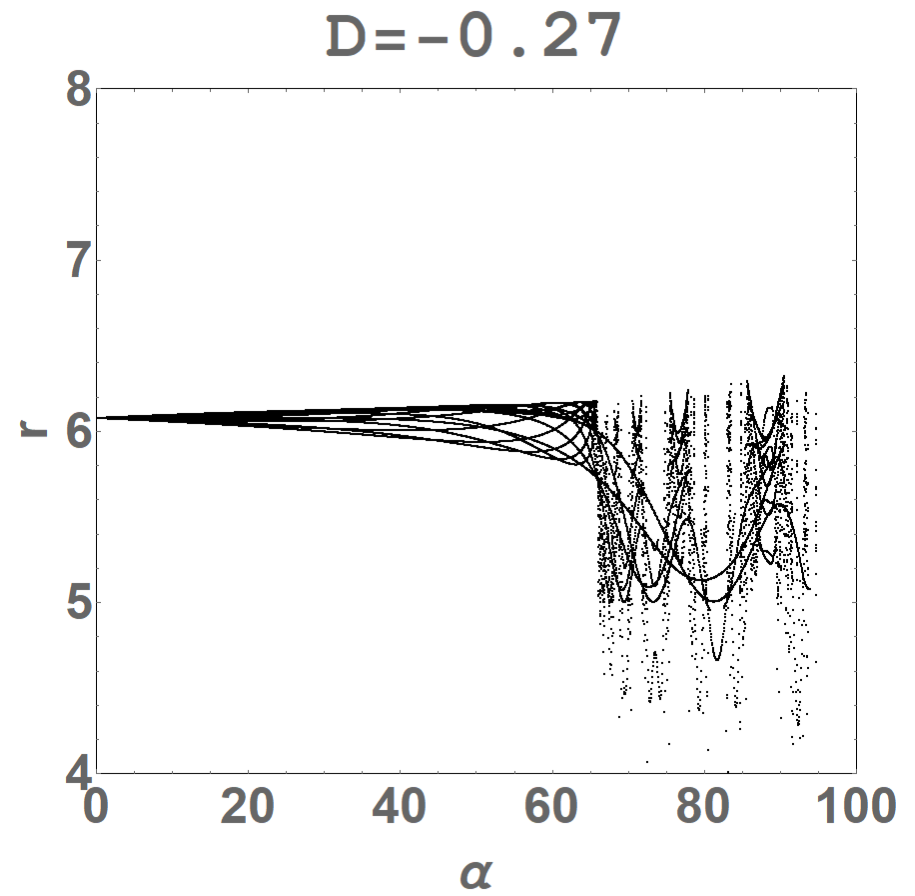}
\includegraphics[width=5.3cm ]{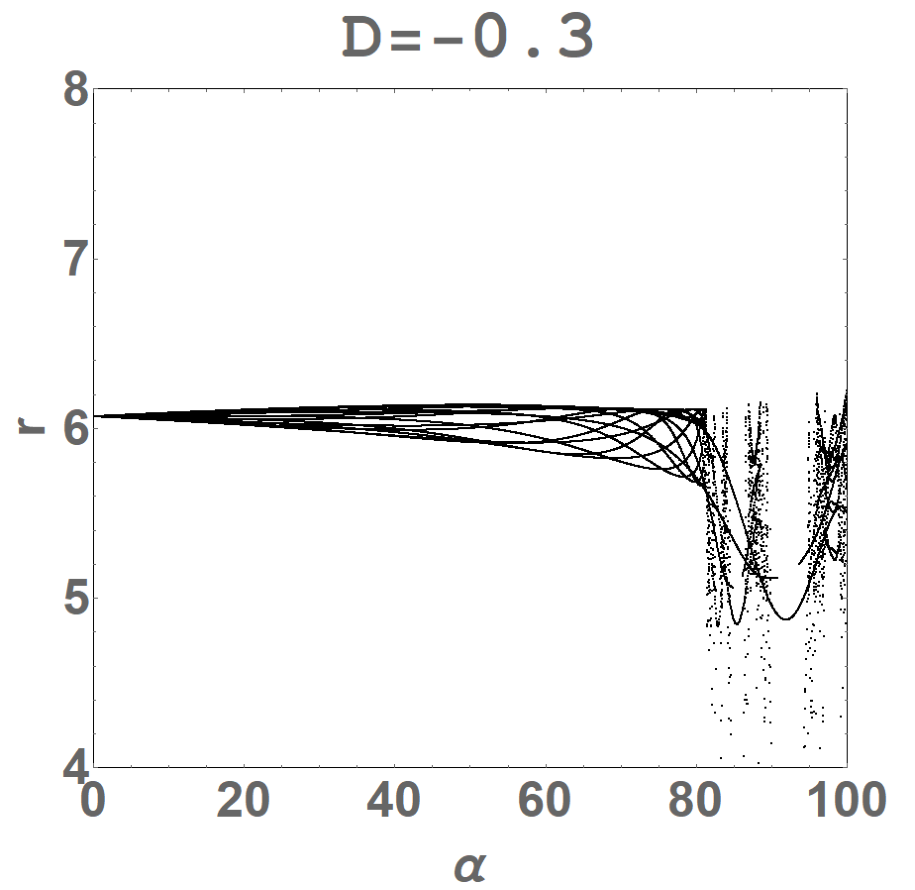}
\includegraphics[width=5.3cm ]{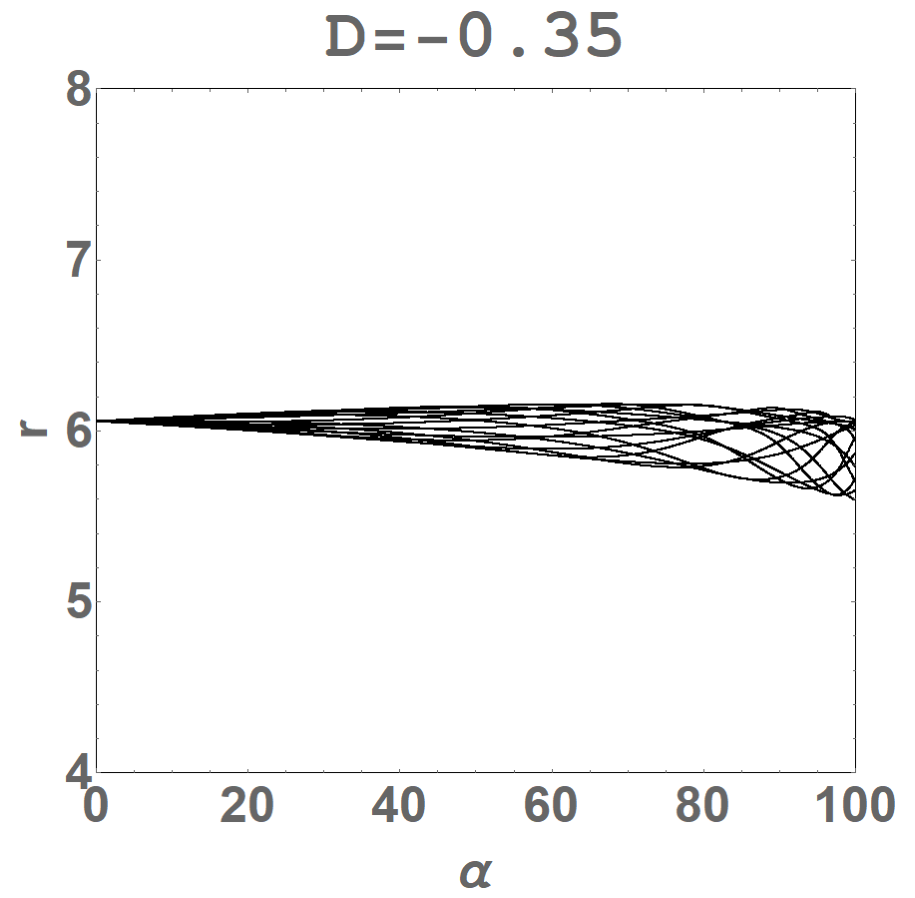}
\caption{The bifurcation changes with the CS coupling parameter $\alpha$ for the fixed spin parameter $a=0.31$ and different dilaton parameters $D$.}\label{fig7}
\end{figure}

\begin{figure}[htbp!]
\includegraphics[width=5.3cm ]{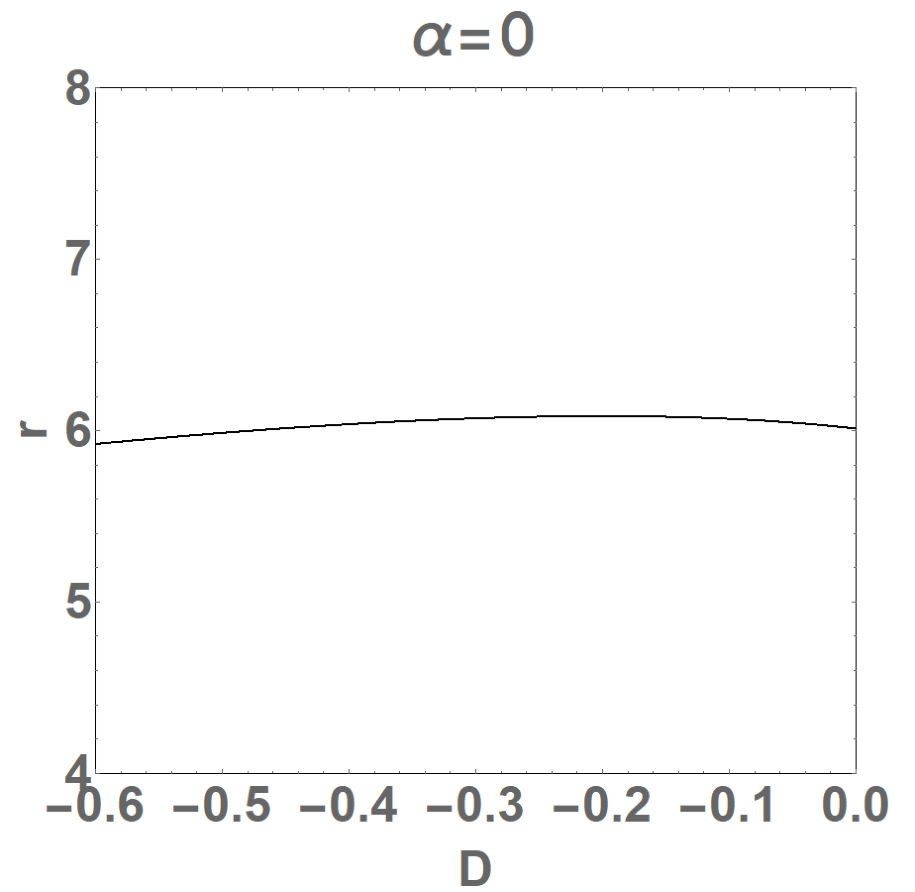}
\includegraphics[width=5.3cm ]{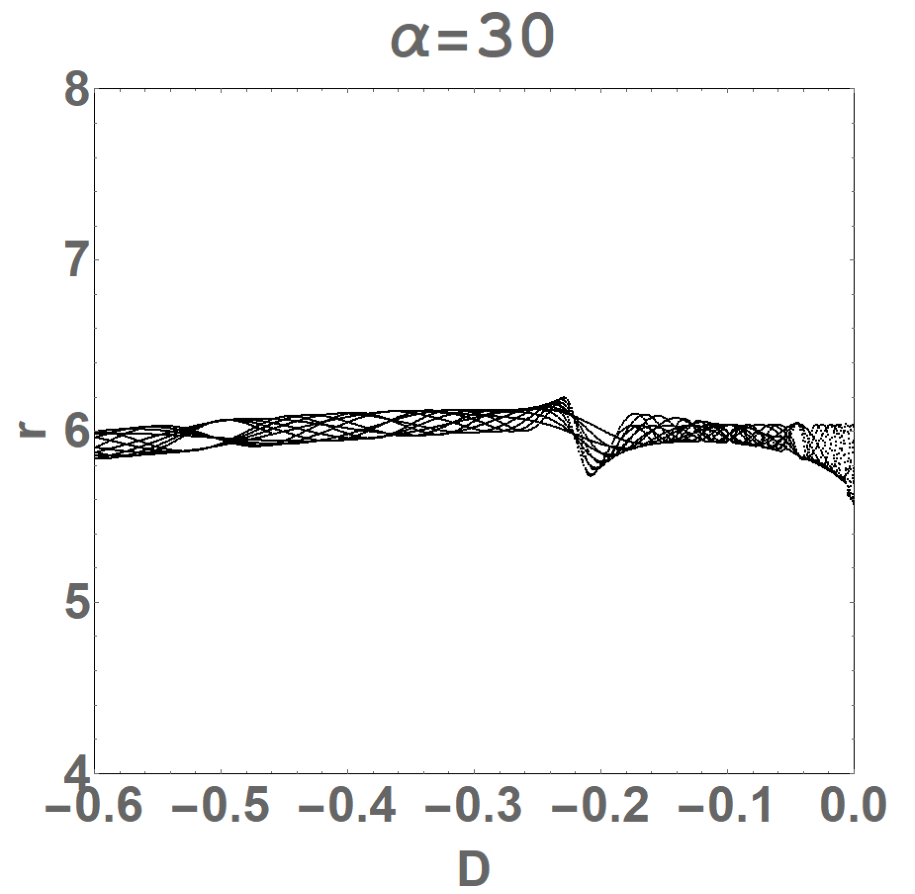}
\includegraphics[width=5.3cm ]{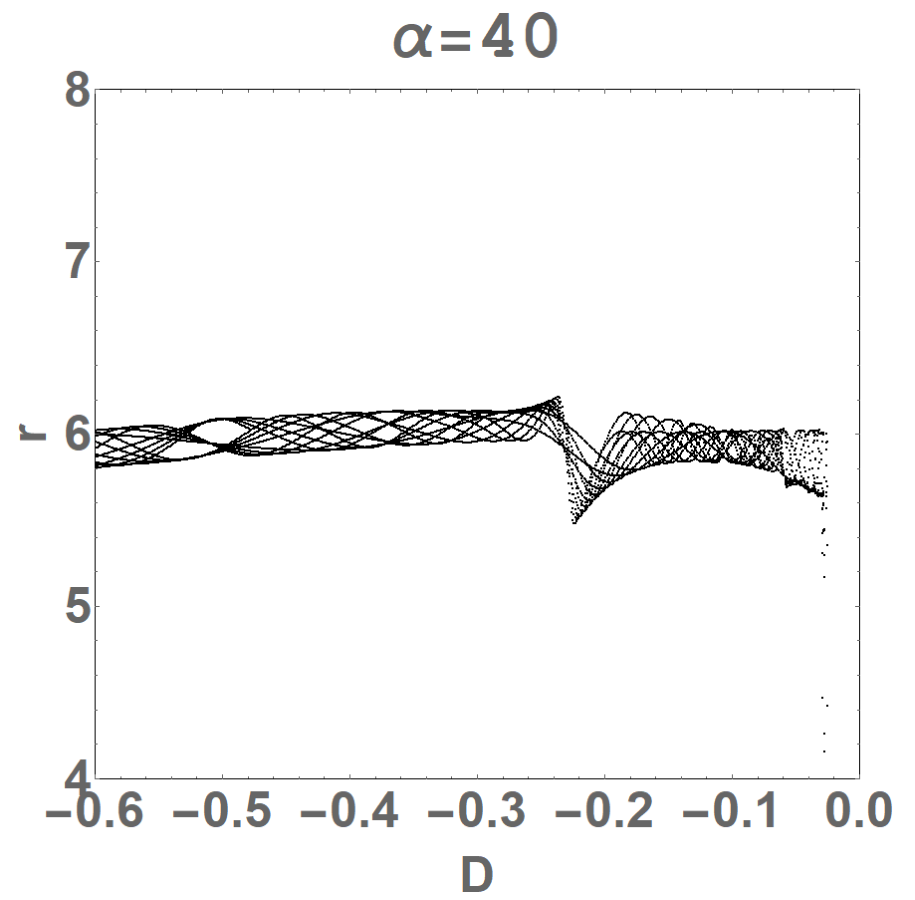}
\includegraphics[width=5.3cm ]{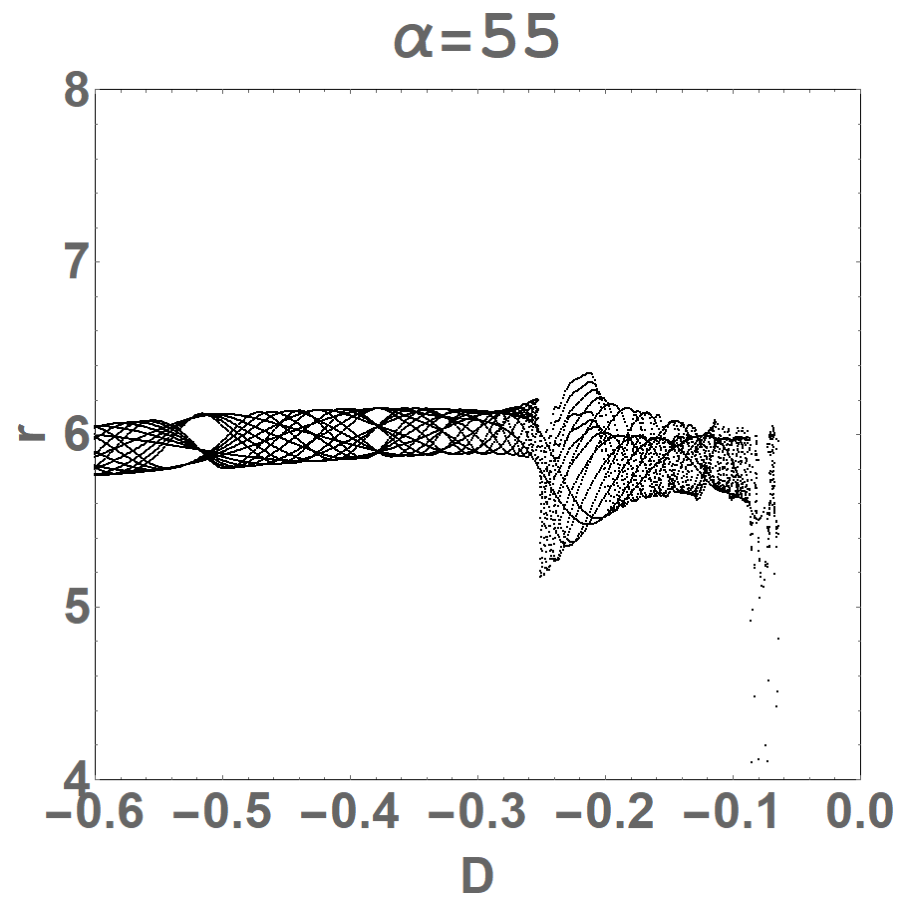}
\includegraphics[width=5.3cm ]{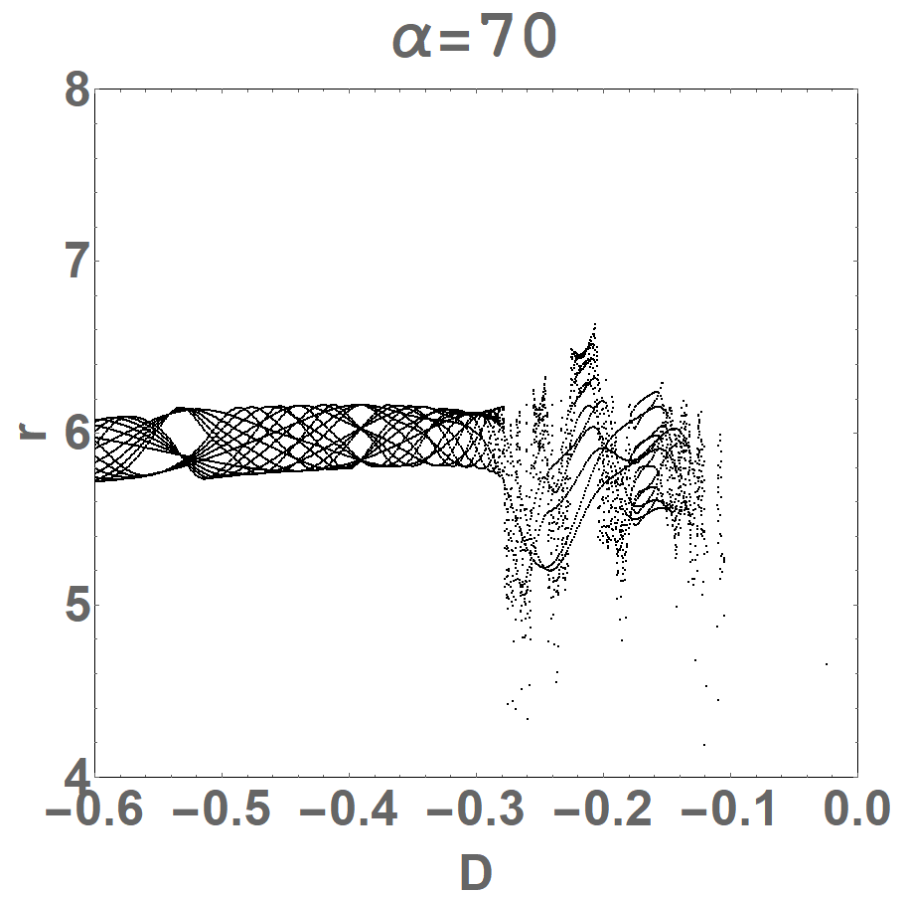}
\includegraphics[width=5.3cm ]{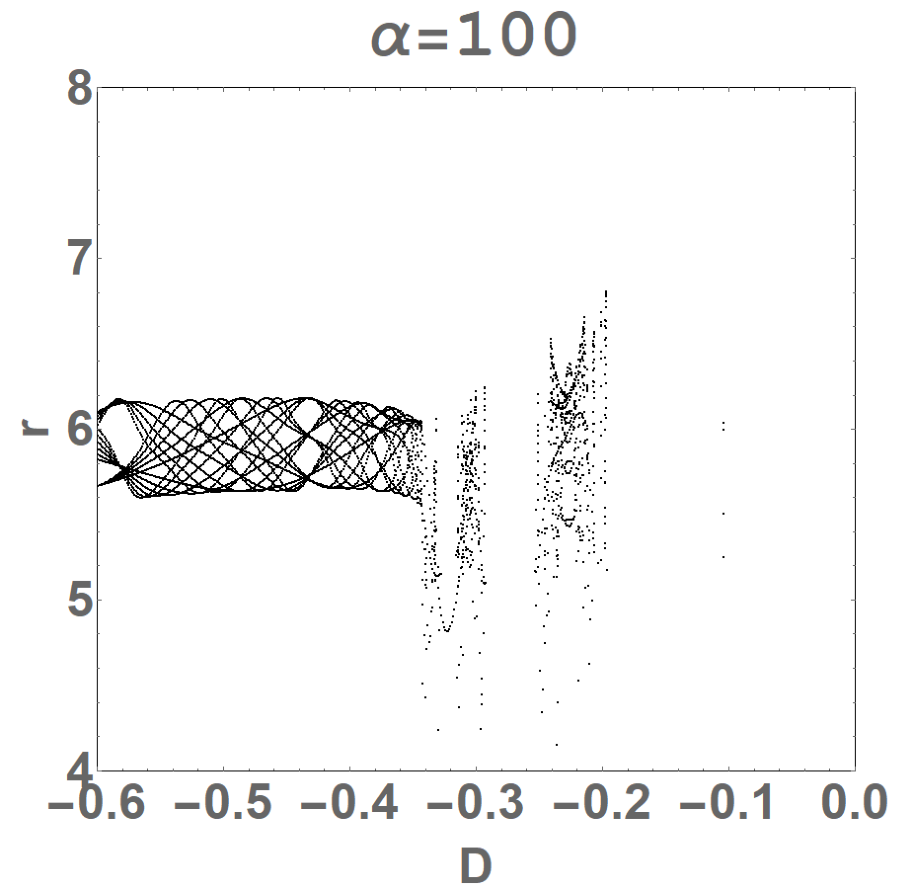}
\caption{The bifurcation changes with the dilaton parameter $D$ for the fixed spin parameter $a=0.31$ and different values of the CS coupling parameter $\alpha$.} \label{fig8}
\end{figure}

\begin{figure}[htbp!]
\includegraphics[width=5.3cm ]{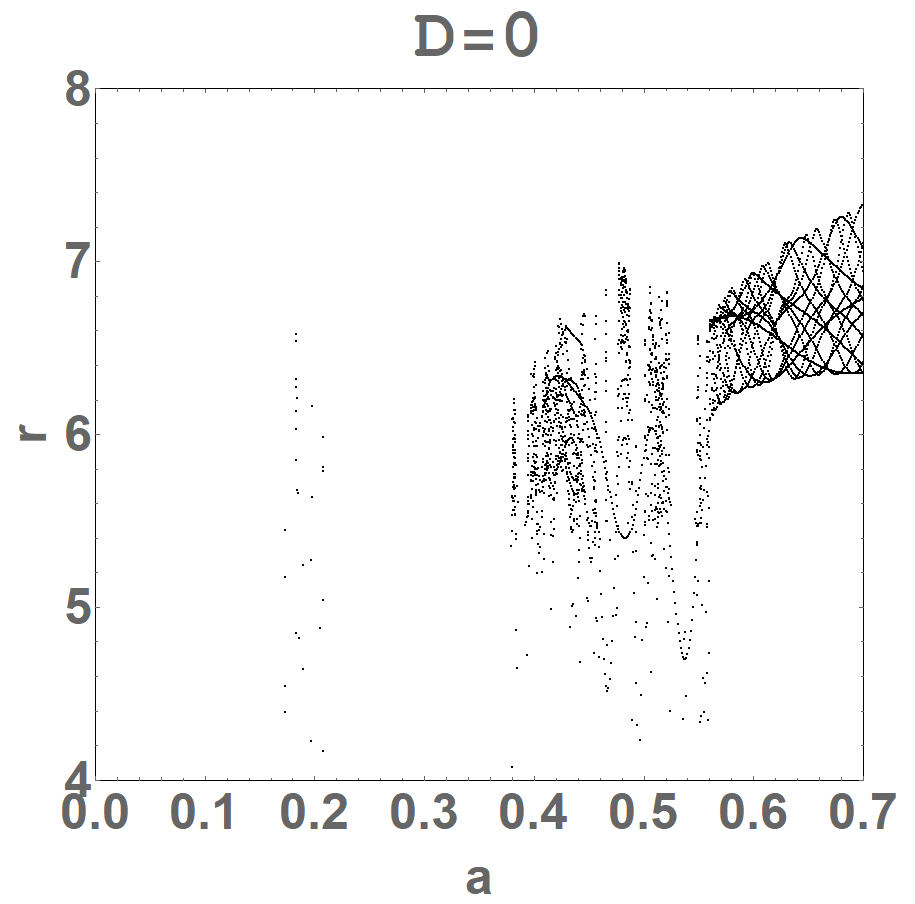}
\includegraphics[width=5.3cm ]{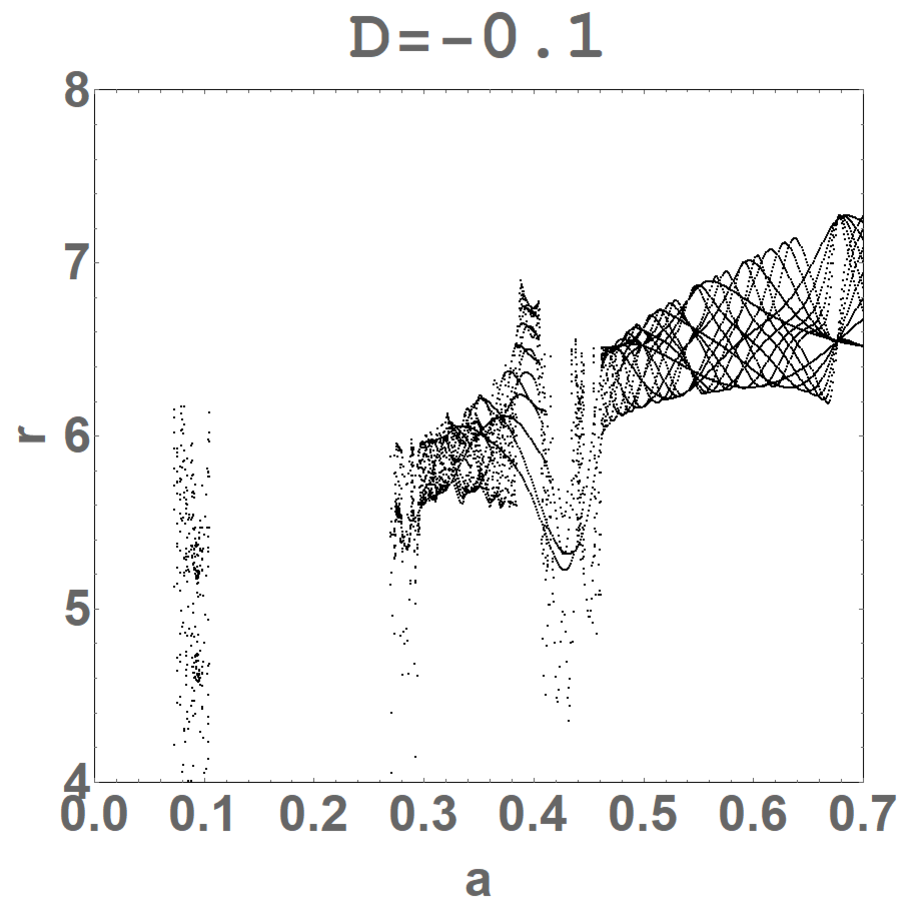}
\includegraphics[width=5.3cm ]{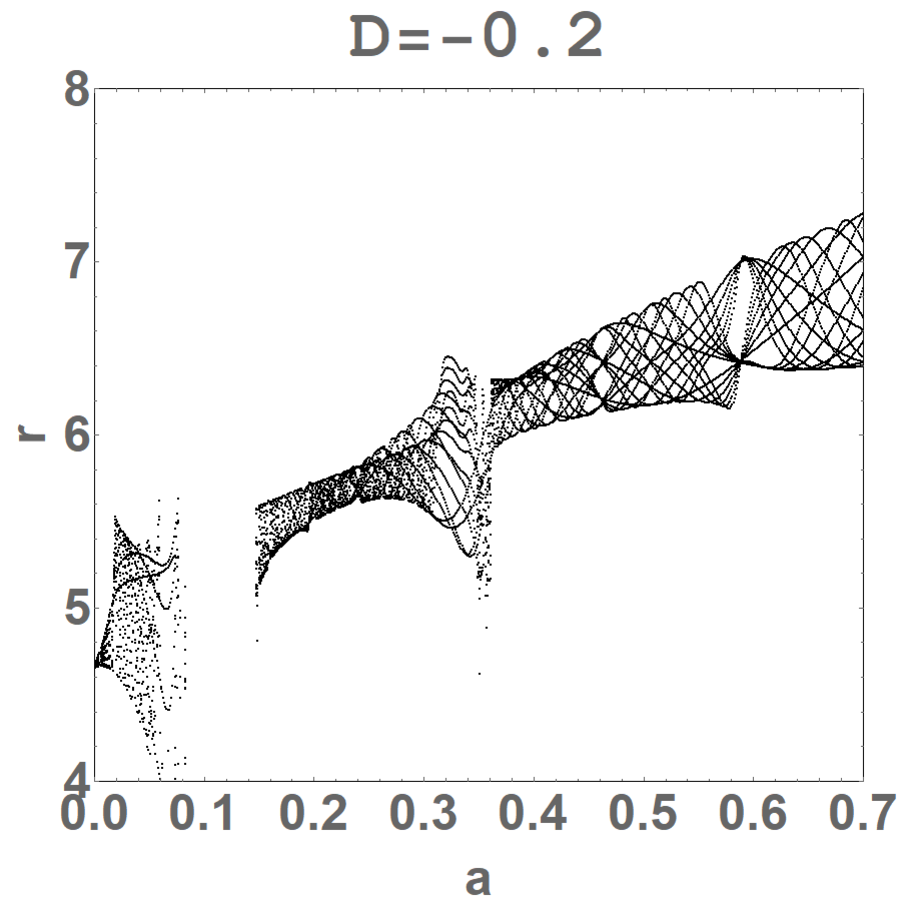}
\includegraphics[width=5.3cm ]{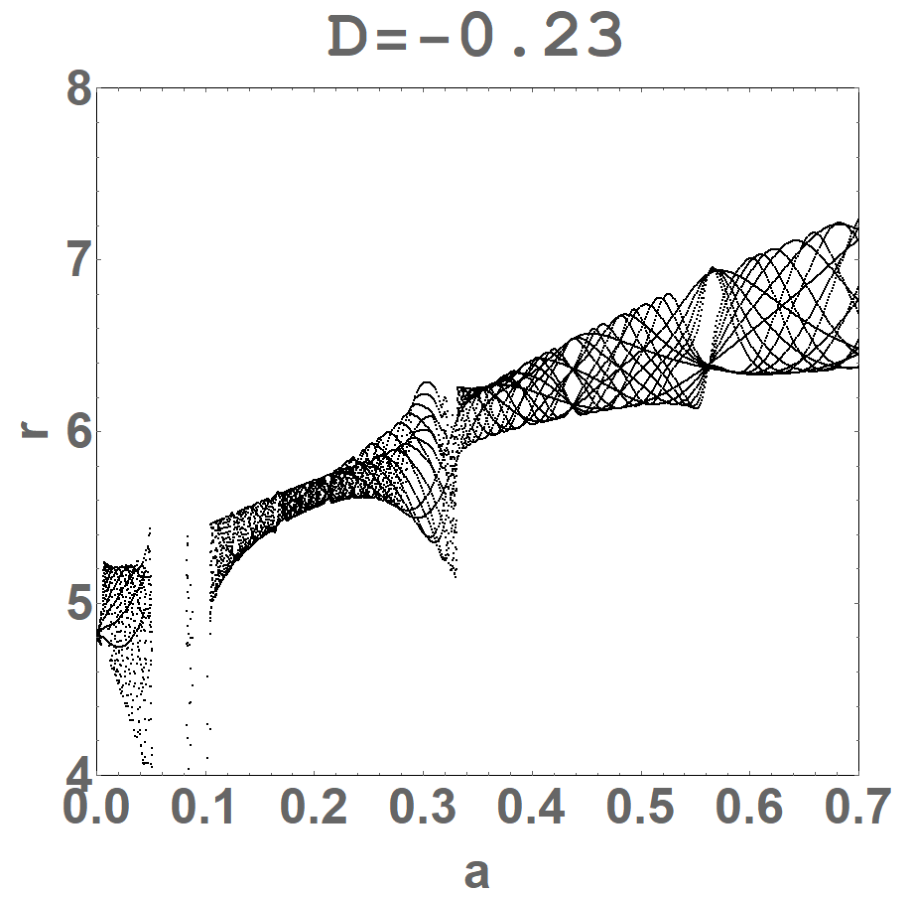}
\includegraphics[width=5.3cm ]{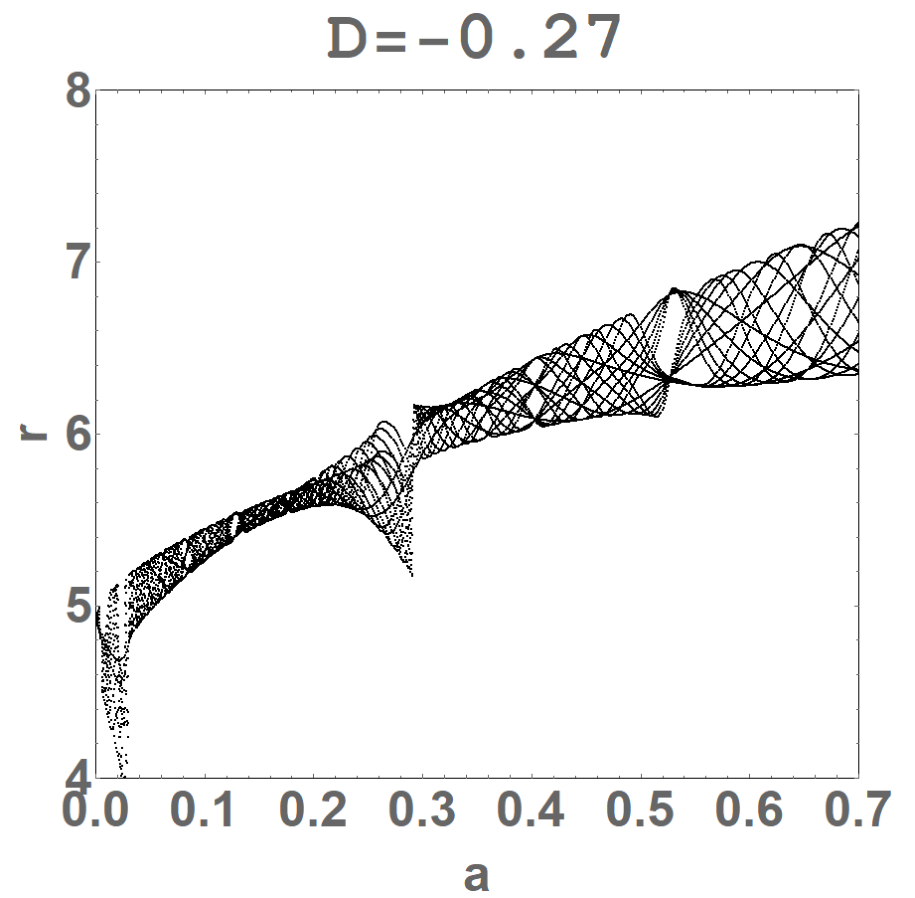}
\includegraphics[width=5.3cm ]{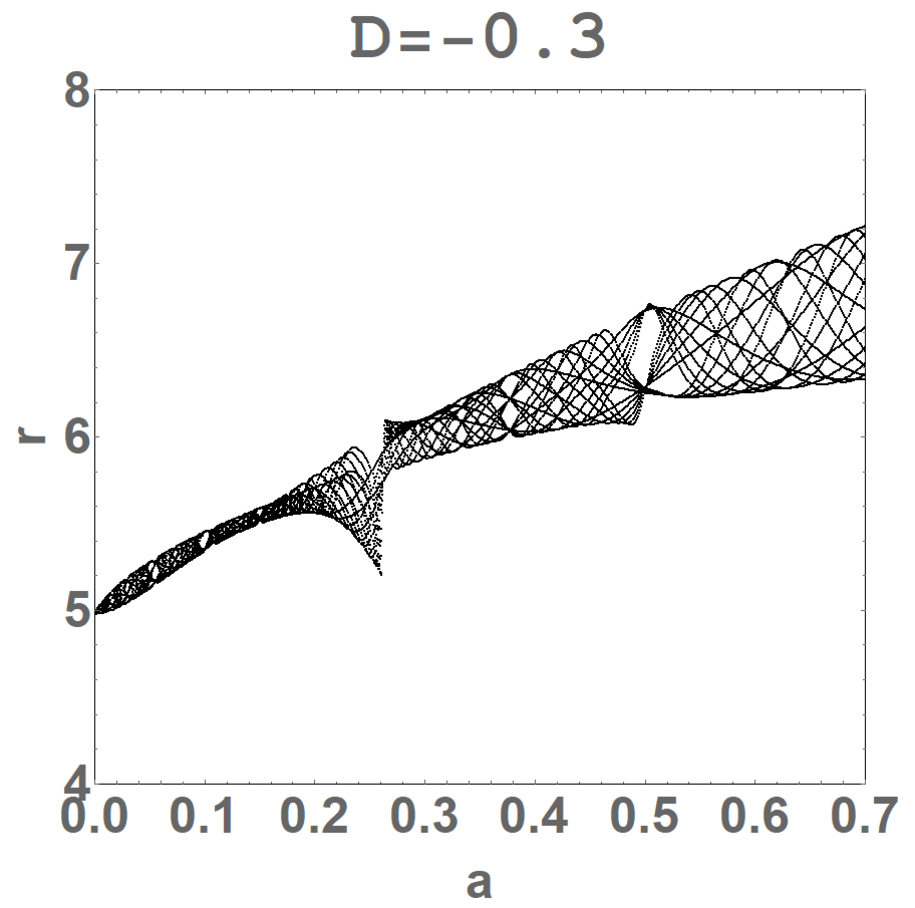}
\caption{The bifurcation changes with the spin parameter $a$ for the fixed CS coupling parameter $\alpha=55$ and different values of the dilaton parameter $D$.}\label{fig9}
\end{figure}

The bifurcation diagram can illustrate the dependence of dynamical behaviors of the particle motion on system parameters.
In Figs. \ref{fig6}-\ref{fig9}, we plot the bifurcation diagram of the radial coordinate of the particle in the EMDA black hole spacetime, and probe effects of the CS coupling parameter $\alpha$, the spin parameter $a$, and the dilaton parameter $D$ on the motion of particles. When $\alpha = 0$ or $a = 0$, one can find that there is no bifurcation for dynamical systems and the motion of the scalar particle is regular in both cases. For the fixed dilaton parameter $D=-0.23$,  with the increase of the spin parameter $a$ of the black hole, Fig. \ref{fig6} shows that the range of $\alpha$ where the chaos occurs first increases and then decreases,  and the corresponding lower limit of $\alpha$ first decreases and then increases. For the fixed spin parameter $a=0.31$, with the increasing $|D|$, Fig. \ref{fig7} shows that the range of $\alpha$ where the chaos appears decreases and the lower limit of $\alpha$ increases. With the increase of $\alpha$, Fig. \ref{fig8} illustrates that the range of $D$ where the chaos occurs increases and the corresponding lower limit of $D$ decreases. For the fixed $\alpha=55$, with the increase of $|D|$,  Fig. \ref{fig9} shows that the range of $a$ in which the chaos appears decreases and the corresponding upper limit of $a$ decreases. These indicate that the motions of the coupled scalar particles heavily depend on the coupling parameter $\alpha$, the black hole parameters $a$ and $D$. Therefore, under the interaction with the CS invariant, the dynamical behavior becomes much richer in a usual rotating EMDA  black hole spacetime.

\begin{figure}[htbp!]
\includegraphics[width=6.5cm ]{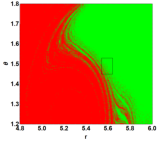}
\includegraphics[width=6.6cm ]{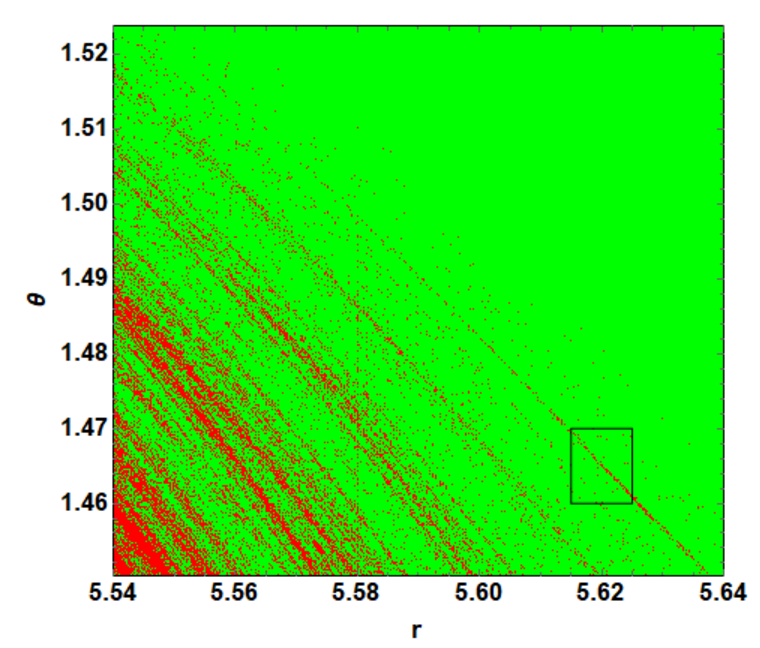}
\includegraphics[width=6.7cm ]{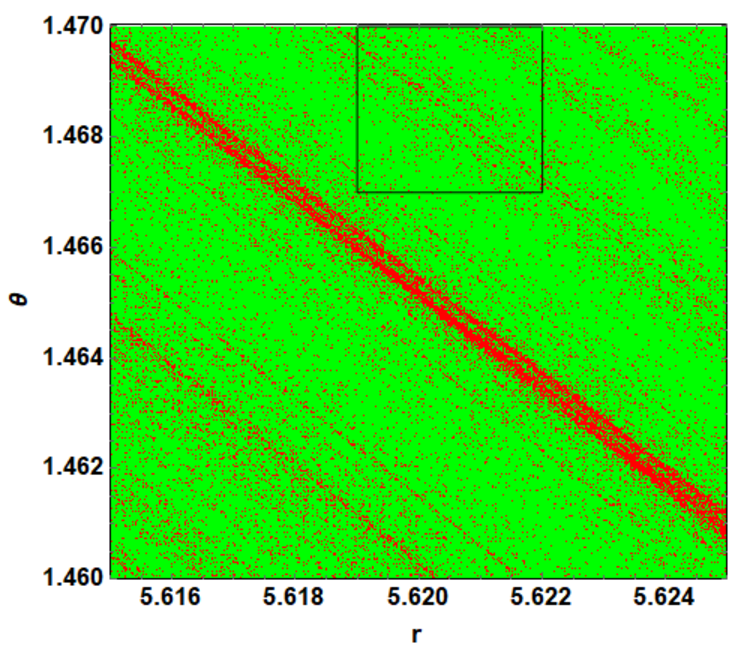}
\includegraphics[width=6.5cm ]{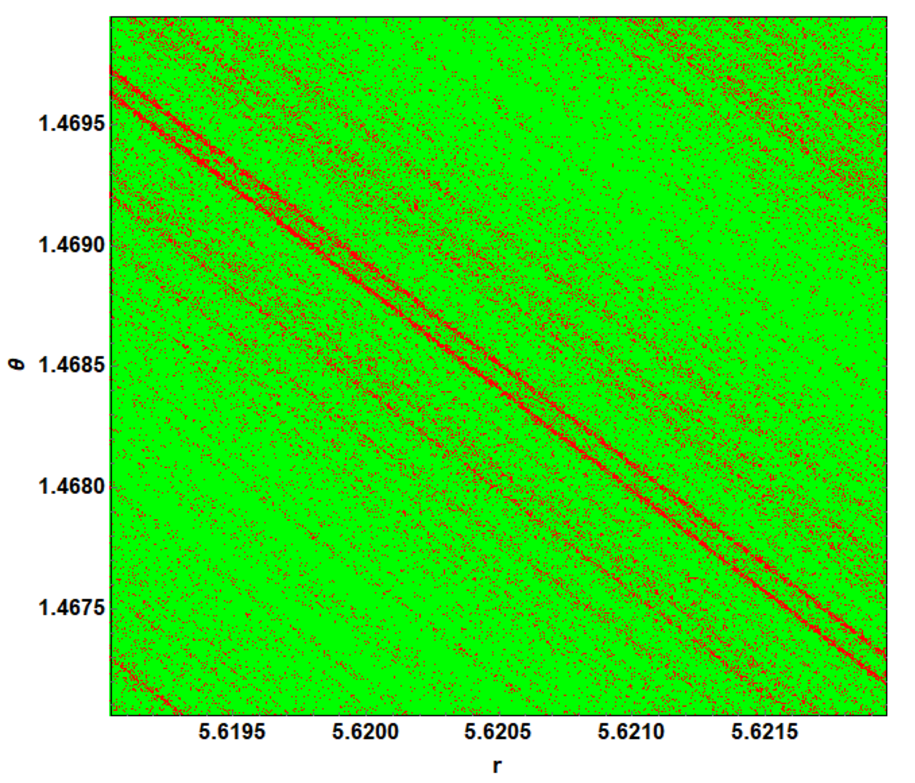}
\caption{The fractal basins of attraction for the coupled scalar particle in a stationary axisymmetric EMDA black hole spacetime with the fixed parameters $D=-0.23$, $a=0.31$, $\alpha=55$, $M_{ADM}=1$, $M=M_{ADM}+D$, $E=0.95$ and $L=3.05M$. }\label{fig10}
\end{figure}

Analysing the basin boundaries of attractors \cite{Cornish,Frolov,Basin} can help to identify the signatures of chaos because the boundary between basins can be fractal as the chaos is present. In Fig. \ref{fig10}, we plot the basins of attraction in a large subset of phase space for a coupled scalar particle in a stationary axisymmetric EMDA black hole spacetime with the fixed parameters $\alpha=55$, $ a=0.31$, $ D=-0.23$, $M_{ADM}=1$, $E=0.95$ and $L= 3.05M$. The initial conditions corresponding to the points shown in the figure, are set to $\dot{r}=0$, and then $\dot{\theta}$ is given by the constraint (\ref{Hcon}), i.e., $h=0$.
The red point corresponds to the particle falling into the black hole along geodesics. The blue point represents the particle escaping into infinity, and the green point denotes the particle oscillating  around the black hole. Here, the condition for the captured particle is set to be $r\leq r_{+}$ and the condition for the escape is to be $r\geq 100r_{+}$. For the green dots, we consider trajectories that neither get captured nor escaped to infinity within $100,000$ iterations. In Fig. \ref{fig10}, it is easy to find that there exist some self-similar fractal fine structures in the basins boundaries  attractors, which also means that there exists the chaotic motion for a coupled scalar particle in the stationary axisymmetric EMDA black hole spacetime.

\section{Summary}

We have studied the motion of test scalar particles coupling to the CS invariant in the stationary axisymmetric EMDA black hole spacetime. The presence of the dilation parameter makes the CS invariant more complex, which yields richer dynamical behaviors of the particles. Applying techniques including the Poincar\'{e} section, fast Lyapunov exponent indicator, bifurcation
diagram and basins of attraction, we confirmed the presence of chaos in the motion of scalar particles interacting with the CS invariants in the rotating EMDA black hole spacetime. The effects of the coupling parameter $\alpha$ and the spin parameter $a$  on the test scalar particle are similar to those in the Kerr black hole case. With the increasing the absolute value
of the dilaton parameter $D$, the number of regular orbits in the Poincar\'{e} section first increases and then decreases, and finally increases again. Meanwhile, the number of chaotic orbits first increases and then decreases. Moreover, with the increasing $|D|$, the chaotic strength for the chaotic orbits first increases and then decreases. For the fixed spin parameter ($a=0.31$), with the increase of $\alpha$, we found that the range of $D$ where the chaos occurs increases and the corresponding lower limit of $D$ decreases. With the increasing $|D|$, the range of $\alpha$ for the appearance of the chaos decreases and the lower limit of $\alpha$ increases. For the fixed coupling parameter ($\alpha= 55$), with the increase of $|D|$, we observed that the range of $a$ in which the chaos appears decreases and the corresponding upper limit of $a$ decreases. These indicate that the motions of the coupled scalar particles heavily depend on the coupling parameter $\alpha$, the black hole parameters $a$ and $D$. Therefore, the CS invariant coupling together with the spin and dilaton parameters yields the richer dynamical behavior of the scalar particle in the stationary axisymmetric EMDA black hole spacetime.

\begin{acknowledgments}

This work was supported by the National Key Research and Development Program of China (Grant No. 2020YFC2201400) and National Natural Science Foundation of China (Grant Nos. 12275078, 12275079 and 12035005).

\end{acknowledgments}

\end{document}